\pdfoutput=1
\documentclass[a4paper,12pt]{article}
\usepackage{amsfonts}
\usepackage{mathrsfs}
\usepackage{amsmath}
\usepackage{amssymb}
\usepackage[medium]{titlesec}
\usepackage{bm}
\usepackage{cite}
\usepackage[normalem]{ulem}
\usepackage{extarrows}
\usepackage{slashed}
\usepackage{isodateo}
\usepackage{graphicx}
\usepackage{xcolor}
\usepackage[CJKbookmarks=true, bookmarksnumbered=true,bookmarksopen=true]{hyperref}
 \hypersetup{colorlinks,%
             linkcolor=[rgb]{0,0.2,0.6}, %
             citecolor=[rgb]{0,0.2,0.6}}
\usepackage[hmargin=.7in,vmargin=1.1in]{geometry}
\usepackage{indentfirst}
\newcommand{\FR}[2]{\displaystyle\frac{\,{#1}\,}{#2}}

\newcommand{\n}{\nonumber}

\graphicspath{{fig/}}

\def\bge{\begin{equation}}
\def\ede{\end{equation}}
\def\bga{\begin{aligned}}
\def\eda{\end{aligned}}
\def\bgp{\begin{pmatrix}}
\def\edp{\end{pmatrix}}
\def\bgs{\begin{subequations}}
\def\eds{\end{subequations}}
\newcommand{\order}[1]{\mathcal{O}({#1})}
\def\di{{\mathrm{d}}}

\def\mb{\mathbf}

\def\pd{\partial}
\def\ld{{\mathscr{L}}}
\def\hd{{\mathscr{H}}}

\def\la{\langle}\def\ra{\rangle}

\def\to{\rightarrow}

\def\ii{\mathrm{i}}

\def\al{\alpha}

\def\ga{\gamma}
\def\de{\delta}
\def\ep{\epsilon}

\def\lam{\lambda}

\def\si{\sigma}

\def\Mp{M_{\text{Pl}}}

\def\Re{\mathrm{Re}\,}
\def\Im{\mathrm{Im}\,}

\newcommand{\ob}[1]{\mkern 2mu \overline{\mkern -2mu #1 \mkern -2mu}\mkern 2mu}
\newcommand{\wt}[1]{\mkern 2mu \widetilde{\mkern -2mu #1 \mkern -2mu}\mkern 2mu}
\newcommand{\wh}[1]{\mkern 2mu \widehat{\mkern-2mu#1\mkern-2mu}\mkern 2mu}

\begin{document}

\title{\vspace{-18mm}\Large\textbf{Schwinger-Keldysh Diagrammatics for Primordial Perturbations}}
\author{Xingang Chen$^{a}$\footnote{Email: xingang.chen@cfa.harvard.edu}{},~~ Yi Wang$^{b}$\footnote{Email: phyw@ust.hk}{},~ and~ Zhong-Zhi Xianyu$^{c,d}$\footnote{Email: xianyu@cmsa.fas.harvard.edu}\\[2mm]
\normalsize{$^a$\emph{Institute for Theory and Computation, Harvard-Smithsonian Center for Astrophysics,}}\\
\normalsize\emph{{60 Garden Street, Cambridge, MA 02138, USA}}\\
\normalsize{$^b$\emph{Department of Physics, The Hong Kong University of Science and Technology,}}\\
\normalsize{\emph{Clear Water Bay, Kowloon, Hong Kong, P.R.China}}\\
\normalsize{$^c$\emph{Center of Mathematical Sciences and Applications, Harvard University,}} \\
\normalsize{\emph{20 Garden Street, Cambridge, MA 02138, USA}}\\
\normalsize{$^d$~\emph{Department of Physics, Harvard University, 17 Oxford Street, Cambridge, MA 02138, USA}}}

\date{}

\maketitle

\begin{abstract}

We present a systematic introduction to the diagrammatic method for practical calculations in inflationary cosmology, based on Schwinger-Keldysh path integral formalism. We show in particular that the diagrammatic rules can be derived directly from a classical Lagrangian even in the presence of derivative couplings. Furthermore, we use a quasi-single-field inflation model as an example to show how this formalism, combined with the trick of mixed propagator, can significantly simplify the calculation of some in-in correlation functions. The resulting bispectrum includes the lighter scalar case ($m<3H/2$) that has been previously studied, and the heavier scalar case ($m>3H/2$) that has not been explicitly computed for this model. The latter provides a concrete example of quantum primordial standard clocks, in which the clock signals can be observably large.

\end{abstract}
\newpage
\tableofcontents

\newpage
\section{Introduction}

In inflationary cosmology, the correlation functions of fields at late times are linked to the observed correlations of primordial fluctuation, and thus are of central importance. These correlation functions are expectation values of field operators at a later time with the initial state (or ``in'' state) given. This is different from the $S$-matrix elements familiar to particle physicists, where both the ``in'' state and the ``out'' state are specified. To calculate such expectation values, there is a well-developed canonical in-in formalism in the literature, which resembles the operator approach to $S$-matrix. In this formalism, one can do calculation by expanding the Dyson series and doing field contractions order by order. See \cite{Chen:2010xka,Wang:2013eqj} for reviews of the canonical in-in formalism.

On the other hand, as is well known in particle physics, the most convenient way to write down expressions for $S$-matrix elements in perturbation theory is to use Feynman diagrams. In particular, when the theory is classically defined by a Lagrangian, it is most convenient to derive the Feynman rules directly from the Lagrangian in the language of path integral. In this way, we free ourselves from the trouble of switching to Hamiltonian and doing field contractions by hand. More importantly, this approach allows us to do calculation neatly in gauge theories (including gravity).

Similarly, it is desirable to compute expectation values by drawing Feynman diagrams in inflationary cosmology. The well-known Schwinger-Keldysh (SK) formalism is in place for this purpose. Introduced originally in the seminal papers by Schwinger \cite{Schwinger:1960qe}, Keldysh \cite{Keldysh:1964ud}, and Feynman and Vernon \cite{Feynman:1963fq}, the SK formalism has found wide applications in many branches of theoretical physics, due to ubiquitous need for computing expectation values. See \cite{Landau10,Chou:1984es,Jordan:1986ug,Haehl:2016pec} for reviews of SK formalism from various aspects.

In fact, the diagrammatic approach of the SK formalism is also known in cosmology community, and has been invoked in the literature (e.g.~\cite{Tsamis:1996qq,Tsamis:1996qm,Seery:2007we,vanderMeulen:2007ah,Seery:2008ax,Leblond:2010yq,Chen:2016nrs}). See also \cite{Calzetta:1986ey,Weinberg:2005vy,Prokopec:2010be,Gong:2016qpq} for introductions to this approach in cosmology. However, it seems that the convenience and advantage of diagrammatic calculation are not widely appreciated in previous studies. For one thing, although the diagrams provide a nice way of organizing perturbation series, apparently they do not by themselves bring much simplification for practical calculation. Indeed, the expressions written following diagrammatic rules must be identical to the ones found from the canonical in-in formalism, if the two formalisms are really equivalent.

But this is not the whole story. In this paper, we would like to emphasize the point that the diagrammatic approach based on SK path integral not only provides a convenient organizing principle for the perturbation series, but also brings considerable simplifications to practical calculations. Since the diagrams make the structures of in-in integrals more transparent, they enable us to simplify the calculation dramatically with the help of some almost trivial tricks. In the example we shall show in this paper, the diagrammatic method can make the numerical calculation $\order{10^4}$ times faster than directly applying the canonical in-in formalism. Moreover, the diagrammatic rules can be conveniently got from a Lagrangian, even in the presence of derivative couplings often encountered in cosmological context. This is again advantageous over the canonical in-in formalism, where the derivative couplings make the form of interaction-picture Hamiltonian rather complicated.

Therefore, in order to make our point clearer and the paper more accessible to a wider range of readers, we feel it appropriate to present a self-contained introduction to SK formalism for inflationary cosmology (Sec.\;\ref{sec_SKF}). In doing so, we also give a careful treatment of derivative couplings which are somewhat overlooked in previous studies using SK formalism in cosmology. In particular, we prove in Appendix \ref{sec_DC} that the Lagrangian used in path integral agrees with the classical Lagrangian to 4th order in power of fields when there are arbitrary derivative couplings. We further illustrate the agreement between canonical in-in formalism and the SK path integral in a specific example involving derivative coupling, in Appendix \ref{sec_SampleDC}.

A far more interesting application of this formalism is presented in Sec.\;\ref{sec_QSFI}, where we calculate the 3-point and 4-point correlation functions (bispectrum and trispectrum, respectively) in quasi-single-field inflation \cite{Chen:2009we,Chen:2009zp, Baumann:2011nk,Assassi:2012zq,Chen:2012ge,Noumi:2012vr,Arkani-Hamed:2015bza,Gong:2013sma,Emami:2013lma,Kehagias:2015jha, Chen:2015lza,Bonga:2015urq,Lee:2016vti}. Together with a small trick of ``mixed propagator'', the diagrammatic method enables us to write down neat expressions for the 3-point and 4-point functions. In the case of 3-point function, the expression is much more compact than what we get from canonical in-in formalism, which also speeds up the numerical calculation significantly as mentioned above. We recover the result of the lighter scalar field case ($m<3H/2$) that has been studied previously \cite{Chen:2009we,Chen:2009zp}, and also provide new results for the heavier scalar field case ($m>3H/2$). The results for 4-point functions shown in this paper are also new.

Concluding remarks, including the discussion about the applicability of this formalism are presented in Sec.\;\ref{sec_Discussion}. Readers who wish a quick look at the diagrammatic rules may go directly to Sec.\;\ref{sec_summary} for a summary.

\section{Schwinger-Keldysh Formalism and Diagrammatic Rules}
\label{sec_SKF}

\subsection{The Set-up}
\label{sec_Setup}

In this section, we present a self-contained introduction to Schwinger-Keldysh formalism and the related diagrammatic rules adapted for primordial perturbations.  Most of this section is a review of known results, except that we provide an explicit treatment of $\ii\ep$-prescription in the inflation background, and also a perturbative proof of the equivalence between the classical Lagrangian and the ``effective'' Lagrangian used in path integral, to 4th order in the power of fields, in the presence of derivative coupling, as elaborated in Appendix \ref{sec_DC}. Furthermore, we come up with a set of diagrammatic notations which is convenient for our purpose.

To be concrete, we begin with a general homogeneous, isotropic, and spatially flat spacetime, described by the FRW metric,
\bge
  \di s^2=-\di t^2+a^2(t)\di \mb x^2,
\ede
where $\di\mb x$ is the distance element in flat time slice of $3$ dimensions. We assume this form of metric only for notational simplicity. The formalism developed here can be generalized to the cases where the conditions such as isotropy and spatial flatness are absent, and also to general $D$ dimensions when dimensional regularization is required. Throughout the paper, we work with conformal time $\tau$, defined through $\di\tau^2=a^2(t)\di t^2$, so that the metric is conformally flat,
\bge
\label{metricCF}
  \di s^2=a^2(\tau)\big(-\di \tau^2+\di\mb x^2\big).
\ede
Suppose we are interested in a field theory described by a classical Lagrangian $\ld_\text{cl}[\phi]$, which is a functional of several dynamical field variables $\phi^A$ where the superscript ``$A$'' labels different fields. In general, the classical equations of motion $\de\ld_\text{cl}[\phi]/\de\phi^A=0$ in FRW background have solutions $\phi^A=\ob\phi^A(\tau)$ with nontrivial time dependence, among which the FRW metric itself is a time-dependent solution of the Einstein equation. On the other hand, in most cases we are interested in solutions being constant over every given time slice $\Sigma_\tau$. Such solutions break spontaneously the time translation symmetry (if there is any), but not the 3-dimensional translation and rotations.

To quantize the theory, we first split the fields $\phi^A(\tau,\mb x)=\ob\phi^A(\tau)+\varphi^A(\tau,\mb x)$ into classical background $\ob\phi^A(\tau)$ and fluctuations $\varphi^A(\tau,\mb x)$. Substituting this back into the Lagrangian $\ld_\text{cl}[\phi]$, we get a new Lagrangian $\ld_\text{cl}[\varphi;\ob\phi(\tau)]$, which can be treated as a functional of fluctuations $\varphi^A(\tau,\mb x)$ starting from the quadratic order. The background solutions $\ob\phi^A(\tau)$, being classical $c$-numbers, can be viewed as time-dependent parameters in the Lagrangian. Therefore, we are effectively considering a Lagrangian $\ld_\text{cl}[\varphi;\tau]$ with field variables $\varphi^A(\tau,\mb x)$ and various time-dependent coupling parameters. In particular, we will also treat the background metric (\ref{metricCF}) as a time-dependent parameter, and therefore we do not spell it out explicitly in various formulae. For example, we write the action as
\bge
  S=\int\di\tau\di^3\mb x\,\ld_\text{cl}[\varphi;\tau],
\ede
with the factor $\sqrt{-g}=a^4(\tau)$ considered as a time-dependent parameter in the Lagrangian. From now on, we shall suppress the $\tau$ dependence in the Lagrangian and simply write $\ld_\text{cl}[\varphi]$ for notational clarity. But it should be kept in mind that the coefficients in the Lagrangian, and also in the Hamiltonian, can have nontrivial time dependence, due to the time-dependent background solutions.

Now we are in the position to quantize the theory. For simplicity, we assume that all of the quantum fluctuation fields $\varphi^A$ are dynamical variables, i.e. there is no constrained variable or gauge redundancy. Such complications are inessential for our presentation of SK formalism, and can be treated on the same footing as ordinary field theory. With this assumption, we can define the canonical conjugate momentum $\pi_A$ and the Hamiltonian as usual,
\begin{align}
\pi_A=&~\FR{\pd\ld_\text{cl}[\varphi]}{\pd\varphi'^A},\\
\label{hd}
\hd[\pi,\varphi]=&~\pi_A\varphi'^A-\ld_\text{cl}[\varphi],
\end{align}
where a prime denotes the derivative with respect to the conformal time, $\varphi'\equiv \pd\varphi/\pd\tau$.

At this stage it is helpful to spell out some details in more concrete examples. Let's consider a Lagrangian $\ld_\text{cl}[\varphi]$ without higher order derivative couplings, which means that $\ld_\text{cl}$ depends on $\varphi'^A$ at most quadratically, \emph{and} that the coefficients of quadratic terms of $\varphi'^A$ are field-independent. We can parameterize such a Lagrangian as,
\bge
\label{ld_nd}
  \ld_\text{cl}=\FR{1}{2}\mathcal{U}_{AB}\varphi'^A\varphi'^B+\mathcal{V}_A(\varphi)\varphi'^A+\mathcal{W}(\varphi),
\ede
where $\mathcal{U}_{AB}$ is a positive definite matrix independent of $\varphi^A$,  while $\mathcal{V}_A(\varphi)$ and $\mathcal{W}(\varphi)$ are arbitrary functions of $\varphi^A$ and its spatial derivatives, but we do not spell out their dependence on spatial derivatives simply for convenience. We can think of $\mathcal{U}_{AB}$ as a metric and use it to lower the field indices, and use its inverse $\mathcal{U}^{AB}=(\mathcal{U}_{AB})^{-1}$ to raise indices. As we shall see below, this is the most general case where we can derive the Lagrangian version of path integral without perturbative expansion.\footnote{In principle, when the ``metric'' $\mathcal{U}_{AB}$ depends on $\varphi^A$, it is also possible to derive a Lagrangian version of path integral without perturbation expansion, but at the expense of an additional annoying field-dependent factor $\det \mathcal{U}$, which still needs a perturbative treatment in diagrammatic calculations. See \cite{WeinbergQFTI} for a discussion. In this paper, we prefer to separate any field dependence in the ``metric'' $\mathcal{U}_{AB}$ into another term and treat it perturbatively. See Appendix \ref{sec_DC} for details.} Then, from the above Lagrangian, we can compute the canonical conjugate momentum and the Hamiltonian,
\begin{align}
  \pi_A=&~\mathcal{U}_{AB}\varphi'^B+\mathcal{V}_A(\varphi),\\
  \label{hd_nd}
  \hd[\pi,\varphi]=&~\FR{1}{2}\pi_A\pi^A-\mathcal{V}_A\pi^A+\FR{1}{2}\mathcal{V}_A\mathcal{V}^A-\mathcal{W}.
\end{align}

As a concrete example of above formalism, here we collect the mode functions and related formulae for a set of massive scalar fields $\varphi_a$ in inflationary (de Sitter) background, which will be used in various places in the following. In this case, we can choose the scale factor $a(\tau)\simeq 1/(-H\tau)$ with $H$ the Hubble parameter, and choose the following Lagrangian,
\begin{align}
\label{Lfree}
\ld_\text{cl}[\varphi]=\sum_a\Big[\FR{1}{2}a^2(\tau)\varphi_a'^2(\tau,\mb x)-\FR{1}{2}a^2(\tau)\big[\pd_i\varphi_a(\tau,\mb x)\big]^2-\FR{1}{2}a^4(\tau)M_a^2\varphi_a^2(\tau,\mb x)\Big]+\cdots,
\end{align}
where $\cdots$ represents interactions, and we have set the masses of scalars $\varphi_a$ to be $M_a$, and assumed all scalar fields have unit sound speed. Then we can represent the field $\varphi_a$ in terms of creation and annihilation operators,
\bge
\label{phimode}
  \varphi_a(\tau,\mb x)=\int\FR{\di^3\mb k}{(2\pi)^3}\Big[u_a(\tau,\mb k)b_a(\mb k)+u_a^*(\tau,-\mb k)b_a^\dag(-\mb k)\Big]e^{\ii\mb k\cdot\mb x},
\ede
where the creation and annihilation operators satisfy the usual commutation relations, of which the only nonvanishing one is,
\bge
  [b_a(\mb k_1),b^\dag_b(\mb k_2)]=(2\pi)^3\de^{(3)}(\mb k_1-\mb k_2)\de_{ab}.
\ede
We can represent the conjugate momentum $\pi_a=a^2(\tau)\varphi_a'$ similarly, as,
\bge
\label{pimode}
  \pi_a(\tau,\mb x)=a^2(\tau)\int\FR{\di^3\mb k}{(2\pi)^3}\Big[u_a'(\tau,\mb k)b_a(\mb k)+u_a^*{}'(\tau,-\mb k)b_a^\dag(-\mb k)\Big]e^{\ii\mb k\cdot\mb x}.
\ede
The mode functions $u_a(\tau,\mb k)$ satisfies the equation of motion derived from $\de\ld_\text{cl}[\varphi]/\de\varphi=0$, which reads,
\bge
  u_a''(\tau,\mb k)-\FR{2}{\tau}u_a'(\tau,\mb k)+\Big(\mb k^2+\FR{M_a^2}{H^2\tau^2}\Big)u_a(\tau,\mb k)=0.
\ede
Assuming the usual Bunch-Davis vacuum as the initial condition, and also the normalization condition,
\bge
  a^2(\tau)\big[u_a(\tau,\mb k)u_a^*{}'(\tau,-\mb k)-u_a'(\tau,\mb k)u_a^*(\tau,-\mb k)\big]=\ii,
\ede
the mode function can be solved to be,
\begin{align}
  u_a(\tau,\mb k)=-\FR{\ii\sqrt{\pi}}{2}e^{\ii\pi(\nu/2+1/4)}H(-\tau)^{3/2}\text{H}_{\nu_a}^{(1)}(-k\tau),
\end{align}
where $\nu_a\equiv\sqrt{9/4-M_a^2/H^2}$, $k\equiv |\mb k|$, and $\text{H}_\nu^{(1)}(z)$ is Hankel function of the first kind. It is useful to note that the mode function reduces to the massless case in the asymptotic past $\tau\to-\infty$, that is,
\bge
\label{Upast}
  u(\tau,\mb k)\to \FR{\ii H\tau}{\sqrt{2k}}e^{-\ii k\tau}.
\ede
It can also be readily checked that the equal-time canonical commutation relation is satisfied,
\bge
  [\varphi_a(\tau,\mb x),\pi_b(\tau,\mb y)]=\ii\de^{(3)}(\mb x-\mb y)\de_{ab}.
\ede

\subsection{Schwinger-Keldysh Path Integral}

The quantity we want to calculate is the expectation value $\la Q(\tau)\ra$ where
\bge
Q(\tau)\equiv\mathcal{O}_1(\tau,\mb x_1)\cdots \mathcal{O}_N(\tau,\mb x_N) ~.
\ede
Here $\mathcal{O}_i(\tau,\mb x_i)$ are operators constructed locally from the field operators in the Lagrangian.
These operators are all on the same time slice $\tau$. The expectation value $\la\cdots\ra$ is taken with respect to a state $|\Omega\ra$ which is fixed at some initial time slice $\tau=\tau_0$, and is usually taken to be the vacuum state with respect to time $\tau$ at $\tau=\tau_0$.
In this paper we are mostly interested in the case,
\bge
\label{Qoperator}
Q(\tau)\equiv\varphi^{A_1}(\tau,\mb x_1)\cdots\varphi^{A_N}(\tau,\mb x_N) ~.
\ede

Before introducing the Schwinger-Keldysh path integral, it is useful to review very briefly the way of computing $\la Q\ra$ in canonical in-in formalism, phrased in operator language. More details can be found in \cite{Chen:2010xka,Wang:2013eqj}. According to this formalism, the expectation value $\la Q\ra$ is firstly rewritten in interaction picture, and then is recast into a Dyson series,
\begin{align}
\la Q(\tau)\ra=\big\la\Omega\big|\ob{F}(\tau,\tau_0)Q_I(\tau)F(\tau,\tau_0)\big|\Omega\big\ra,
\end{align}
where $Q_I$ is the operator $Q$ written in interaction picture, $F(\tau,\tau_0)$ is the usual Dyson series, and $\ob{F}(\tau,\tau_0)$ is its Hermitian conjugate, i.e.,
\begin{align}
&F(\tau,\tau_0)={\text{T}}\exp\bigg(-\ii\int_{\tau_0}^\tau\di\tau_1\,H_I(\tau_1)\bigg),\\
&\ob{F}(\tau,\tau_0)=\ob{\text{T}}\,\exp\bigg(\ii\int_{\tau_0}^\tau\di\tau_1\,H_I(\tau_1)\bigg),
\end{align}
where T and $\ob{\text{T}}$ represent time ordering and anti-time ordering, respectively, and $H_I(\tau)$ is the interacting part of the Hamiltonian written in interaction picture,
\begin{align}
\label{Hint}
  &H_I(\tau)=\int\di^3\mb x\,\hd_I[\pi_I,\varphi_I;\tau],
  &&\hd_I[\pi_I,\varphi_I;\tau]\equiv \hd[\pi_I,\varphi_I;\tau]-\hd_0[\pi_I,\varphi_I;\tau].
\end{align}
Here $\hd$ is the full Hamiltonian and $\hd_0$ is its ``free'' part, which is usually just the quadratic terms in $\hd$ though the way of splitting $\hd$ into free part and interacting part is not unique.
The complication here is not only that one needs to do perturbative expansion of Dyson series and the Wick contraction by hand, but also that the coupling terms in interaction-picture Hamiltonian $\hd_I$ are not simply the sign reverse of the corresponding Lagrangian when derivative couplings are present. In fact, there may be non-negligible new terms generated in $\hd_I$ which are absent in the classical Lagrangian $\ld_\text{cl}[\varphi]$. See \cite{Huang:2006eha} and \cite{Wang:2013eqj} for a review. Also see (\ref{Hpert}) and the subsequent discussions in Appendix \ref{sec_DC} of this paper for further details.

Now we are going to write down a path integral representation for $\la Q\ra$ starting from the Heisenberg picture, where the initial state is time independent, and thus the desired quantity is simply given by,
\bge
  \la Q\ra=\la\Omega|Q(\tau)|\Omega\ra.
\ede
In order to derive a path-integral representation of $\la Q\ra$ in a similar way as we do for the $S$-matrix, the first step is to recast $\la Q\ra$ into an amplitude between an in state and an out state. For this purpose, we can naturally identify $|\Omega\ra$ to be the in state, but we do not know an out state. The trick out of this problem, as usual in physics, is to ``sum over our ignorance''. That is, we choose a time slice $\Sigma_f$ at any time $\tau_f\geq\tau$, usually just taken at $\tau_f=\tau$ as we shall assume in the following, and insert a complete basis of states $1=\sum |O_\al\ra\la O_\al|$ on $\Sigma_f$ into $\la Q\ra$,
\begin{align}
\label{Qev}
\la Q\ra=\sum_\al\la\Omega|O_\al\ra \la O_\al | \varphi^{A_1}(\tau,\mb x_1)\cdots\varphi^{A_N}(\tau,\mb x_N)\big|\Omega\ra.
\end{align}
At this moment we immediately realize an ambiguity that we can insert the complete basis at many different places. For now, we stick to the prescription that we insert the complete basis to the left of all operator insertions, and we will come back later to comment on the other choices of inserting the complete basis. (For the special case where all operator insertions sitting on the final time slice $\Sigma_f$ as is considered here, the position of inserting the complete basis turns out to be irrelevant, as we shall see below.)

Now, the two factors on the right hand side of (\ref{Qev}) have the desired form of an $S$-matrix element, and thus are amenable to path integral representation. In particular, the $\la\Omega|O_\al\ra$ resembles the conjugation of a vacuum amplitude, with the time order of the in and out states reversed. For this reason, we call it the anti-time-ordered factor. Similarly, we call $\la O_\al |Q|\Omega\ra$ the time-ordered factor.

To write down path integral representation for both factors, we proceed as usual, by foliating the spacetime between initial slice $\Sigma_0$ and final slice $\Sigma_f$ with infinitely many time slices, and then inserting complete eigenbasis $1=\sum_{\varphi(\tau_i)}|\varphi(\tau_i)\ra\la \varphi(\tau_i)|$ of field operator $\varphi^A$ and the complete eigenbasis $1=\sum_{\pi(\tau_i)}|\pi(\tau_i)\ra\la\pi(\tau_i)|$ of the conjugate momentum $\pi_A$, at each slice $\Sigma_i$. Specifically, the time-ordered factor can be written as a path integral over the field configurations $\varphi_{+}^{A}(\tau,\mb x)$ and the conjugate momentum $\pi_{+A}(\tau,\mb x)$ as,
\begin{align}
\la O_\al | Q_I(\tau) |\Omega \ra
=&\int\mathcal{D}\varphi_{+}\mathcal{D}\pi_{+}\exp\bigg[\ii \int_{\tau_0}^{\tau_f}\di\tau \di^3\mb x\,\Big(\pi_{+A}\varphi_{+}'^{A}-\hd[\pi_+,\varphi_+]\Big)\bigg]\n\\
 &\times  \varphi_+^{A_1}(\tau,\mb x_1)\cdots\varphi_+^{A_N}(\tau,\mb x_N)\big\la O_\al \big|\varphi_+(\tau_f)\big\ra\big\la \varphi_+(\tau_0)\big|\Omega\big\ra,
\end{align}
while the anti-time ordered factor can be written as a path integral over the field configurations $\varphi_-^A(\tau,\mb x)$ and the conjugate momentum $\pi_{-A}(\tau,\mb x)$,
\begin{align}
\la \Omega |O_\al\ra
=&\int\mathcal{D}\varphi_{-}\mathcal{D}\pi_{-}\exp\bigg[-\ii \int_{\tau_0}^{\tau_f}\di\tau \di^3\mb x\,\Big(\pi_{-A} \varphi_{-}'^{A}-\hd[\pi_-,\varphi_-]\Big)\bigg]\n\\
 &\times \big\la \varphi_-(\tau_f) \big| O_\al \big\ra\big\la \Omega \big| \varphi_-(\tau_0)\big\ra.
\end{align}
In above expressions, we introduced variables with $+$ or $-$ indices for time-ordered and anti-time-ordered factors, respectively, and the Hamiltonian $\hd[\pi,\varphi]$ has exactly the same functional dependence on $\pi_\pm$ and $\varphi_\pm$ as the Hamiltonian in Heisenberg picture derived in (\ref{hd}). Combining the two factors and summing over $\al$, we get,
\begin{align}
\label{QevHPI}
\la Q \ra=&~\int\mathcal{D}\varphi_{+}\mathcal{D}\pi_{+}\mathcal{D}\varphi_{-}\mathcal{D}\pi_{-}\,\varphi_+^{A_1}(\tau,\mb x_1)\cdots\varphi_+^{A_N}(\tau,\mb x_N) \n\\
&~\times\exp\bigg[\ii \int_{\tau_0}^{\tau_f}\di\tau \di^3\mb x\,\Big(\pi_{+A}\varphi_{+}'^{A}-\hd[\pi_+,\varphi_+]\Big)\bigg]\n\\
&~\times\exp\bigg[-\ii \int_{\tau_0}^{\tau_f}\di\tau \di^3\mb x\,\Big(\pi_{-A}\varphi_{-}'^{A}-\hd[\pi_-,\varphi_-]\Big)\bigg]\n\\
&~\times\la\Omega|\varphi_-(\tau_0)\ra\la\varphi_+(\tau_0)|\Omega\ra\prod_{A,\mb x}\de\Big(\varphi_{+}^{A}(\tau_f,\mb x)-\varphi_{-}^{A}(\tau_f,\mb x)\Big),
\end{align}
{ where the path integral is unconstrained at both $\tau=\tau_0$ and $\tau=\tau_f$, which means in particular that one should integrate over all possible states $|\varphi_-(\tau_0)\ra$ and $\la\varphi_+(\tau_0)|$ appeared in the integrand.} As a result, we get two copies of path integrals, one goes forward in time, and the other goes backward in time, and the two are sewn at future time limit $\tau_f$ by the condition $\varphi_{+}^{A}(\tau_f)=\varphi_{-}^{A}(\tau_f)$.

The integral over momenta $\pi_{\pm A}$ can be directly carried out for the theory without higher order derivative couplings, namely the one given by the Lagrangian (\ref{ld_nd}). This is because in such cases, the Hamiltonian (\ref{hd_nd}) is quadratic in the momentum, and thus the path integral over momenta in (\ref{QevHPI}) is simply a Gaussian. Therefore, substituting (\ref{hd_nd}) into (\ref{QevHPI}), we get,
\begin{align}
\label{freepiint}
  &\int\mathcal{D}\pi_+\,\exp\bigg[\ii \int_{\tau_0}^{\tau_f}\di\tau \di^3\mb x\,\Big(\pi_{+A}\varphi_{+}'^{A}-\hd[\pi_+,\varphi_+]\Big)\bigg]\n\\
  =&~\exp\bigg[\ii\int_{\tau_0}^{\tau_f}\di\tau \di^3\mb x\,\Big(\FR{1}{2}\mathcal{U}_{AB}\varphi_+'^A\varphi_+'^B+\mathcal{V}_A(\varphi_+)\varphi_+'^A+\mathcal{W}(\varphi_+)\Big)\bigg],
\end{align}
in which the integrand is exactly the classical Lagrangian $\ld_\text{cl}[\varphi_+]$ (\ref{ld_nd}) but written in terms of $\varphi_{+A}$. Similarly, the path integral over $\pi_-$ can be carried out, giving another factor $e^{-\ii\int\di\tau\di^3\mb x\,\ld_{\text{cl}}[\varphi_-]}$. Then, the expectation value (\ref{QevHPI}) is further simplified to,
\begin{align}
\label{QevLPI}
\la Q \ra=&~\int\mathcal{D}\varphi_{+} \mathcal{D}\varphi_{-} \,\varphi_+^{A_1}(\tau,\mb x_1)\cdots\varphi_+^{A_N}(\tau,\mb x_N)
\exp\bigg[\ii\int_{\tau_0}^{\tau_f}\di\tau\di^3\mb x\,\Big(\ld_\text{cl}[\varphi_+]-\ld_{\text{cl}}[\varphi_-]\Big)\bigg]\n\\
&~\times\la\Omega|\varphi_-(\tau_0)\ra\la\varphi_+(\tau_0)|\Omega\ra\prod_{A,\mb x}\de\Big(\varphi_{+}^{A}(\tau_f,\mb x)-\varphi_{-}^{A}(\tau_f,\mb x)\Big).
\end{align}
However, when the theory contains higher order derivative couplings, the momentum path integral cannot be carried out in closed form. We show in Appendix \ref{sec_DC} that the path integral over momentum can still be carried out perturbatively, and the result again agrees with the classical Lagrangian up to 4th order in the power of fields. Therefore, we shall assume from now on that (\ref{QevLPI}) also holds for theories with higher order derivative couplings. This is an important point for cosmological application, because higher order derivative couplings appear frequently in this context.

The expression (\ref{QevLPI}) is almost in the desired form of a path integral, weighted by the exponential of an action $e^{\ii S}$, where $S[\varphi_\pm]=\int\di\tau\di^3\mb x(\ld[\varphi_+]-\ld[\varphi_-])$, except for the second line, where we have three additional factors, two inner products of states and one $\delta$-functional. The meaning of the $\delta$-functional is obvious: it tells us that the path integrals for $\varphi_+^A$ and $\varphi_-^A$ are to be sewn together at the final time slice $\tau=\tau_f$. On the other hand, we show in the next several paragraphs that the meaning of the two inner products $\la\Omega|\varphi_-(\tau_0)\ra$ and $\la\varphi_+(\tau_0)|\Omega\ra$ is to provide the correct $\ii\ep$-prescription for the time integral, following the treatment of \cite{WeinbergQFTI}. The main point is to recognize that the inner products are nothing but the vacuum wave functionals represented in field basis, and therefore, they must satisfy the defining equation for the vacuum $b_A|\Omega\ra=0$, also rewritten in the field basis, where $b_A$ is the annihilation operator.

Therefore, the first step is to represent the annihilation operator $b_A$ in terms of field $\varphi^A$ and its conjugate momentum $\pi_A$. To be concrete, we restrict ourselves temporarily with the case of inflation, assuming the scale factor $a(\tau)\simeq 1/|H\tau|$ where $H$ is the Hubble parameter, and the generalization to other cases should be straightforward. We further consider the Lagrangian (\ref{Lfree}) of several massive scalar fields as an example, where we can take field metric $\mathcal{U}_{AB}=\de_{AB}$. We assume that the interactions are switched off at asymptotic past $\tau=\tau_0\to-\infty$, so that $\varphi^A$ becomes essentially free fields at $\tau_0$. Furthermore, the mode function at past infinity (\ref{Upast}) tells that these fields are not only free but also effectively massless at $\tau_0$, so they have very simple mode functions,
\bge
\label{uModeTau0}
  u_A(\tau_0,\mb k)=\FR{\ii H\tau}{\sqrt{2k}}e^{-\ii k\tau}.
\ede
Then from (\ref{phimode}) and (\ref{pimode}), we can solve the annihilation operator $b_A(\mb k)$ as,
\bge
  b_A(\mb k)=-\ii\int\di^3\mb x\,\Big[a^2(\tau)u_A^*{}'(\tau,-\mb k)\varphi_A(\tau,\mb x)-u_A^*(\tau,-\mb k)\pi_A(\tau,\mb x)\Big]e^{-\ii\mb k\cdot\mb x},~~(\text{no sum over $A$})
\ede
Note further that the conjugate momentum operator $\pi_A(\tau,\mb x)=-\ii\de/\de\varphi^A(\tau,\mb x)$ in the field basis, therefore the equation $b_A|\Omega\ra=0$ represented in the field basis $|\varphi_+(\tau_0)\ra$ has the following form,
\begin{align}
\label{VWFplus}
  0=&\int\di^3\mb x\,e^{-\ii\mb k\cdot\mb x}\Big[\FR{\de}{\de\varphi_{+}^A(\tau_0,\mb x)}-\FR{\ii a^2(\tau_0) u_A^*{}'(\tau_0,-\mb k)}{u_A^*(\tau_0,\mb k)}\varphi_{+A}(\tau_0,\mb x)\Big]\la\varphi_{+}(\tau_0)|\Omega\ra\n\\
  =&\int\di^3\mb x\,e^{-\ii\mb k\cdot\mb x}\Big[\FR{\de}{\de\varphi_{+}^{A}(\tau_0,\mb x)}+a^2(\tau_0)k\varphi_{+A}(\tau_0,\mb x)\Big]\la\varphi_+(\tau_0)|\Omega\ra,
\end{align}
where we have used the fact that the mode function $u_A(\tau,\mb k)$ reduces to the massless one (\ref{uModeTau0}) at $\tau_0\to-\infty$. The equation above can be solved by a Gaussian functional,
\begin{align}
\label{InnerProdPlusM}
\la\varphi_+(\tau_0)|\Omega\ra
=&~\mathcal{N}\exp\bigg[-\FR{1}{2}\int\di^3\mb x\di^3\mb y\, \mathcal{E}_{AB}(\tau_0;\mb x,\mb y)\varphi_{+}^{A}(\tau_0,\mb x)\varphi_{+}^{B}(\tau_0,\mb y)\bigg]\n\\
=&~\mathcal{N}\exp\bigg[-\FR{\ep}{2}\int_{\tau_0}^{\tau_f}\di\tau\int\di^3\mb x\di^3\mb y\, \mathcal{E}_{AB}(\tau;\mb x,\mb y)\varphi_+^A(\tau,\mb x)\varphi_+^B(\tau,\mb y)e^{\ep\tau}\bigg].
\end{align}
Here $\ep$ is an infinitesimal positive parameter, $\mathcal{N}$ is the normalization factor for the wave functional, and $\mathcal{E}_{AB}(\tau;\mb x,\mb y)$ can be further solved by substituting (\ref{InnerProdPlusM}) back into (\ref{VWFplus}), with the following result,
\bge
  \mathcal{E}_{AB}(\tau;\mb x,\mb y)=a^2(\tau)\int\FR{\di^3\mb k}{(2\pi)^3}e^{\ii\mb k\cdot(\mb x-\mb y)}k\de_{AB}.
\ede
Substituting this result back to (\ref{InnerProdPlusM}), we get,
\bge
\label{InnerProdPlus}
\la\varphi_+(\tau_0)|\Omega\ra
=\mathcal{N}\exp\bigg[-\FR{\ep}{2}\int_{\tau_0}^{\tau_f}\di\tau\,a^2(\tau)\int\FR{\di^3\mb k}{(2\pi)^3}\,k\varphi_{+A}(\tau,\mb k)\varphi_+^A(\tau,-\mb k)\bigg],
\ede
where we have neglected the factor $e^{\ep\tau}$, which is a correction of higher order in $\ep$. Similarly, the other inner product $\la\Omega|\varphi_-(\tau_0)\ra$ in (\ref{QevLPI}) can be represented by,
\begin{align}
\label{InnerProdMinus}
\la\Omega|\varphi_-(\tau_0)\ra=\mathcal{N}^*\exp\bigg[-\FR{\ep}{2}\int_{\tau_0}^{\tau_f}\di\tau\,a^2(\tau)\int\FR{\di^3\mb k}{(2\pi)^3}\,k\varphi_{-A}(\tau,\mb k)\varphi_-^A(\tau,-\mb k)\bigg].
\end{align}
The factors $\mathcal{N}$ and $\mathcal{N^*}$ are unimportant because they also appear in $\la 1\ra=1$ and thus are canceled out if we divide $\la Q\ra$ by $\la 1\ra=1$.(\footnote{ It should be noted that the Gaussian initial state obtained here is correct initial state only for free field theory. The correct vacuum state for the corresponding interacting theory could develop some non-Gaussian components relative to the Gaussian state which can be treated perturbatively in powers of couplings. This can be important for calculation of loop diagrams but we shall not consider it further as we are mainly focus on tree-level diagrams in this paper.})

Now, we are ready to substitute (\ref{InnerProdPlus}) and (\ref{InnerProdMinus}) back into (\ref{QevLPI}), which amounts to introducing two additional terms into the Lagrangian, as follows,
\bge
  \ld_\text{cl}[\varphi_\pm]\to\ld_\text{cl}[\varphi_\pm] \pm\FR{\ii\ep}{2}\int\di\tau\,a^2(\tau)\int\FR{\di^3\mb k}{(2\pi)^3} k\varphi_{\pm A}(\mb k)\varphi_\pm^A(-\mb k).
\ede
Comparing this expression with (\ref{Lfree}), we see that the additional terms amount to a correction $k\tau\to (1-\ii\ep)k\tau$ in the mode function (\ref{uModeTau0}) for time-ordered variable $\varphi_+^A$, and a correction $k\tau\to (1+\ii\ep)k\tau$ in the mode function for anti-time-ordered variable $\varphi_-^A$, which are further equivalent to a tiny deformation of the time direction into the complex plane, in such a way that $\tau\to(1-\ii\ep)\tau$ for time-ordered part and $\tau\to(1+\ii\ep)\tau$ for anti-time-ordered part. Therefore, we have shown that the net effect of the two inner products is to provide the correct $\ii\ep$-prescription for the path integral, and from now on, we shall always assume that the time integral has been deformed appropriately, and thus take the two inner products away from (\ref{QevLPI}). Then, we are left with the following expression,
\begin{align}
\label{LPIDelta}
\la Q \ra=&~\int\mathcal{D}\varphi_{+} \mathcal{D}\varphi_{-} \,\varphi_+^{A_1}(\tau,\mb x_1)\cdots\varphi_+^{A_N}(\tau,\mb x_N)
\exp\bigg[\ii\int_{\tau_0}^{\tau_f}\di\tau\di^3\mb x\,\Big(\ld_\text{cl}[\varphi_+]-\ld_{\text{cl}}[\varphi_-]\Big)\bigg]\n\\
 &~\times\prod_{A,\mb x}\de\Big(\varphi_{+}^{A}(\tau_f,\mb x)-\varphi_{-}^{A}(\tau_f,\mb x)\Big).
\end{align}
This is the desired SK path integral representation of the expectation value evaluated at a given time. The expression itself is intuitive, which illustrates very well the three steps of writing down the SK path integral, as summarized in the following box.
\begin{center}
\fbox{\begin{minipage}{0.95\textwidth}
\vspace{3mm}
\paragraph{Mnemonic for SK path integral}
\begin{enumerate}
\item Double the fields $\varphi^A$ in the classical Lagrangian as $\varphi^A_\pm$;
\item Assign a Lagrangian $\ld$ to $\varphi_+$, which is identical to the classical Lagrangian even in the presence of derivative couplings, and also assign $-\ld$ to $\varphi_-$;
\item ``Sew'' the path integral for $\varphi_\pm$ at final time slice $\tau=\tau_f$ by the $\de$-functional in (\ref{LPIDelta}).
\end{enumerate}
\vspace{1mm}
\end{minipage}}
\end{center}

Since the path integral (\ref{LPIDelta}) is in quite a standard form, it is routine to derive a set of diagrammatic rules for perturbative calculation, which we shall outline in next subsection. At the end of this subsection, we comment briefly on the ambiguity we mentioned below (\ref{Qev}). From the derivations in this subsection, it is now clear that we are basically free to put field insertions $\varphi^{A_i}(\tau,\mb x_i)$ either in time-ordered factor $\la O_\al|\cdots|\Omega\ra$ or in anti-time-ordered factor $\la\Omega|\cdots|O_\al\ra$. After written into the path-integral form, two choices amount to labelling the field insertions with $+$ index or with $-$ index, respectively. Therefore, depending on the details of the operators, we could have at most $2^N$ different correlation functions for $\langle Q \rangle$ with $Q$ now a product of $\varphi^{A_1}(\tau_1, \mathbf{x}_1), \cdots, \varphi^{A_N}(\tau_N, \mathbf{x}_N)$ with any possible ordering, due to the choice of $\pm$ index for each point. However, it is also clear for the case of equal-time insertions in (\ref{Qoperator}) that if these field insertions are put on the final time slice $\tau_f$, or rather, if the final time slice $\tau_f$ is just chosen at the time of field insertions which is almost always the case, then it makes no difference to choose $+$ or $-$ index for each field insertions, due to the constraint of $\de$-functional in (\ref{LPIDelta}).

\subsection{Generating Functional and Diagrammatic Rules}

As usual, the expectation value (\ref{LPIDelta}) can be computed from a generating functional by taking functional derivative, and the procedure can be summarized neatly into a set of diagrammatic rules. This is the usual story of Feynman diagrams, so we only outline the main steps, and refer the readers to any standard textbook of quantum field theory for details. For notational simplicity, we shall consider only one real scalar field $\varphi$ in this subsection, and the generalization to many scalars should be straightforward. We shall also comment on generalization to fields with nonzero spin afterwards.

To begin with, we introduce external sources $J_\pm(\tau,\mb x)$ for the scalar fields $\varphi_\pm(\tau,\mb x)$, and define the generating functional $Z[J_+,J_-]$ as,
\begin{align}
\label{GF}
  Z[J_+,J_-]=\int\mathcal{D}\varphi_{+} \mathcal{D}\varphi_{-} \,
\exp\bigg[\ii\int_{\tau_0}^{\tau_f}\di\tau\di^3\mb x\,\Big(\ld_\text{cl}[\varphi_+]-\ld_\text{cl}[\varphi_-]+J_+\varphi_+-J_-\varphi_-\Big)\bigg].
\end{align}
Then, a general amplitude $\la \varphi_{a_1}(\tau,\mb x_1)\cdots\varphi_{a_N}(\tau,\mb x_N)\ra$ $(a_1,\cdots,a_N =\pm)$ can be calculated by taking functional derivative as usual,
{
\begin{align}
\label{QevZ}
&~\la \varphi_{a_1}(\tau,\mb x_1)\cdots\varphi_{a_N}(\tau,\mb x_N)\ra\n\\
=&\int\mathcal{D}\varphi_{+} \mathcal{D}\varphi_{-} \,\varphi_{a_1}(\tau,\mb x_1)\cdots\varphi_{a_N}(\tau,\mb x_N)
\exp\bigg[\ii\int_{\tau_0}^{\tau_f}\di\tau\di^3\mb x\,\Big(\ld_\text{cl}[\varphi_+]-\ld_\text{cl}[\varphi_-]+J_+\varphi_+-J_-\varphi_-\Big)\bigg]\n\\
=&~\FR{\de}{\ii a_1\de J_{a_1}(\tau,\mb x_1)}\cdots\FR{\de}{\ii a_N\de J_{a_N}(\tau,\mb x_N)}Z[J_+,J_-]\bigg|_{J_\pm=0}.
\end{align}
}%
We can do perturbative calculation of this amplitude as usual, by splitting the Lagrangian $\ld_\text{cl}$ into the ``free'' part $\ld_0$ and ``interaction'' part $\ld_\text{int}$,
\bge
  \ld_\text{cl}[\varphi]=\ld_0[\varphi]+\ld_\text{int}[\varphi].
\ede
Very often, we take the ``free'' part $\ld_0$ to include all terms quadratic in $\varphi^A$ (note that we have assumed that $\ld_\text{cl}$ starts from quadratic order), and call the rest of terms ``interactions'' and put them in $\ld_\text{int}$. But sometimes it is also helpful to view two-point mixing among different fields as interactions, of which we show an example in the next section.

Then, we can rewrite the generating functional (\ref{GF}) as,
\begin{align}
  Z[J_+,J_-]=&~\exp\bigg[\ii\int_{\tau_0}^{\tau_f}\di\tau\di^3\mb x\,\Big(\ld_\text{int}\Big[\FR{\de}{\ii\de J_+}\Big]-\ld_\text{int}\Big[-\FR{\de}{\ii\de J_-}\Big]\Big)\bigg]Z_0[J_+,J_-],\\
  Z_0[J_+,J_-]\equiv&\int\mathcal{D}\varphi_+\mathcal{D}\varphi_-\,\exp\bigg[\ii\int_{\tau_0}^{\tau_f}\di\tau\di^3\mb x\,\Big(\ld_0[\varphi_+]-\ld_0[\varphi_-]+J_+\varphi_+-J_-\varphi_-\Big)\bigg],
\end{align}
where the path integral $Z_0[J_+,J_-]$ is simply a Gaussian and thus can be carried out explicitly. Then one can expand the first line perturbatively to desired order, and combine it with (\ref{QevZ}) to calculate the expectation value $\la Q\ra$. This procedure generates the diagrammatic rules we are seeking for. Therefore, we stop our very general discussions at this point, and turn to examples to show how the diagrammatic rules are derived and how they work.

\subsubsection{Propagators}

Firstly, let us work out the tree-level propagators, which are defined to be the following two-point functions,
\begin{align}
-\ii \Delta_{ab}(\tau_1,\mb x_1;\tau_2,\mb x_2)=\FR{\de}{\ii a \de J_{a}(\tau_1,\mb x_1)}\FR{\de}{\ii b \de J_{b}(\tau_2,\mb x_2)}Z_0[J_+,J_-]\bigg|_{J_{\pm}=0},
\end{align}
where $a,b=\pm$. Due to the different choices for $a,b$ indices, we have 4 types of propagators. For example, the $(++)$-type propagator can be worked out as,
\bgs
\label{PropX}
\begin{align}
-\ii \Delta_{++}(\tau_1,\mb x_1;\tau_2,\mb x_2)
=&~\FR{\de}{\ii \de J_+(\tau_1,\mb x_1)}\FR{\de}{\ii \de J_+(\tau_2,\mb x_2)}Z_0[J_+,J_-]\bigg|_{J_{\pm}=0}\n\\
=&~\int\mathcal{D}\varphi_+\mathcal{D}\varphi_-\,\varphi_+(\tau_1,\mb x_1)\varphi_+(\tau_2,\mb x_2)e^{\ii\int\di\tau\di^3\mb x\,(\ld_0[\varphi_+]-\ld_0[\varphi_-])}\n\\
=&\sum_{\al}\la\Omega|O_\al\ra\la O_\al|\text{T}\{\varphi(\tau_1,\mb x_1)\varphi(\tau_2,\mb x_2)\}|\Omega\ra\n\\
=&~\la\Omega|\text{T}\{\varphi(\tau_1,\mb x_1)\varphi(\tau_2,\mb x_2)\}|\Omega\ra.
\end{align}
Similarly, we can work out the rest of three propagators. The $(--)$-type is given by,
\begin{align}
-\ii \Delta_{--}(\tau_1,\mb x_1;\tau_2,\mb x_2)
=&~\FR{-\de}{\ii \de J_-(\tau_1,\mb x_1)}\FR{-\de}{\ii \de J_-(\tau_2,\mb x_2)}Z_0[J_+,J_-]\bigg|_{J_{\pm}=0}\n\\
=&~\int\mathcal{D}\varphi_+\mathcal{D}\varphi_-\,\varphi_-(\tau_1,\mb x_1)\varphi_-(\tau_2,\mb x_2)e^{\ii\int\di\tau\di^3\mb x\,(\ld_0[\varphi_+]-\ld_0[\varphi_-] )}\n\\
=&\sum_{\al}\la\Omega|\ob{\text{T}}\{\varphi(\tau_1,\mb x_1)\varphi(\tau_2,\mb x_2)\}|O_\al\ra\la O_\al|\Omega\ra\n\\
=&~\la\Omega|\ob{\text{T}}\{\varphi(\tau_1,\mb x_1)\varphi(\tau_2,\mb x_2)\}|\Omega\ra.
\end{align}
The $(+-)$-type,
\begin{align}
-\ii \Delta_{+-}(\tau_1,\mb x_1;\tau_2,\mb x_2)
=&~\FR{\de}{\ii \de J_+(\tau_1,\mb x_1)}\FR{-\de}{\ii \de J_-(\tau_2,\mb x_2)}Z_0[J_+,J_-]\bigg|_{J_{\pm}=0}\n\\
=&~\int\mathcal{D}\varphi_+\mathcal{D}\varphi_-\,\varphi_+(\tau_1,\mb x_1)\varphi_-(\tau_2,\mb x_2)e^{\ii\int\di\tau\di^3\mb x\,(\ld_0[\varphi_+]-\ld_0[\varphi_-] )}\n\\
=&\sum_{\al}\la\Omega|\varphi(\tau_2,\mb x_2)|O_\al\ra\la O_\al|\varphi(\tau_1,\mb x_1)|\Omega\ra\n\\
=&~\la\Omega|\varphi(\tau_2,\mb x_2)\varphi(\tau_1,\mb x_1)|\Omega\ra.
\end{align}
Finally, the $(-+)$-type,
\begin{align}
-\ii \Delta_{-+}(\tau_1,\mb x_1;\tau_2,\mb x_2)
=&~\FR{-\de}{\ii \de J_-(\tau_1,\mb x_1)}\FR{\de}{\ii \de J_+(\tau_2,\mb x_2)}Z_0[J_+,J_-]\bigg|_{J_{\pm}=0}\n\\
=&~\int\mathcal{D}\varphi_+\mathcal{D}\varphi_-\,\varphi_-(\tau_1,\mb x_1)\varphi_+(\tau_2,\mb x_2)e^{\ii\int\di\tau\di^3\mb x\,(\ld_0[\varphi_+]-\ld_0[\varphi_-] )}\n\\
=&\sum_{\al}\la\Omega|\varphi(\tau_1,\mb x_1)|O_\al\ra\la O_\al|\varphi(\tau_2,\mb x_2)|\Omega\ra\n\\
=&~\la\Omega|\varphi(\tau_1,\mb x_1)\varphi(\tau_2,\mb x_2)|\Omega\ra.
\end{align}
\eds

In practical calculations, we usually go to the 3-momentum space, thanks to the translational and rotational symmetries on each time slice. In this case, we can express the field $\varphi$ in terms of mode functions $u(\tau,\mb k)$ and creation/annihilation operators of given 3-momentum $\mb k$, as we summarized at the end of Sec.~\ref{sec_Setup}, and then substitute the result back into (\ref{PropX}). In this way, we get the propagators in momentum space, which are related to their coordinate-space counterparts via,
\begin{align}
G_{ab}(k;\tau_1,\tau_2)=-\ii\int\di^3\mb x\, e^{-\ii\mb k\cdot  \mb x }\Delta_{ab}(\tau_1,\mb x;\tau_2,\mb 0),
\end{align}
where we have defined the awkward factor $-\ii$ into $G_{ab}$, so that it does not appear in the momentum-space diagrammatic rules. Furthermore, we have written the momentum dependence in $G_{ab}$ as $k=|\mb k|$ because the propagator only depends on the magnitude of the 3-momentum but not its direction, due to the rotational symmetry. Then, the tree-level propagators in the 3-momentum space can be easily worked out to be,
\bgs
\label{SKprop}
\begin{align}
G_{++}(k;\tau_1,\tau_2)=&~G_>(k;\tau_1,\tau_2)\theta(\tau_1-\tau_2)+G_<(k;\tau_1,\tau_2)\theta(\tau_2-\tau_1),\\
G_{+-}(k;\tau_1,\tau_2)=&~G_<(k;\tau_1,\tau_2),\\
G_{-+}(k;\tau_1,\tau_2)=&~G_>(k;\tau_1,\tau_2),\\
G_{--}(k;\tau_1,\tau_2)=&~G_<(k;\tau_1,\tau_2)\theta(\tau_1-\tau_2)+G_>(k;\tau_1,\tau_2)\theta(\tau_2-\tau_1),
\end{align}
\eds
where
\bgs
\label{Ggs}
\begin{align}
G_>(k;\tau_1,\tau_2)\equiv&~u(\tau_1,k)u^*(\tau_2,k),\\
G_<(k;\tau_1,\tau_2)\equiv&~u^{*}(\tau_1,k)u(\tau_2,k).
\end{align}
\eds
It is clear that various propagators are not fully independent. Only three of the four propagators in (\ref{SKprop}) are linearly independent. If we also take account of complex conjugation, then we further have the relations $G_>^*=G_<$, $G_{++}^*=G_{--}$, and $G_{+-}^*=G_{-+}$.

In the literature, the Keldysh basis \cite{Keldysh:1964ud} (or its simple variations) is also widely used as another representation of propagators. In this basis, one defines the following propagators,
\bgs
\label{KBasis}
\begin{align}
  G_A(k;\tau_1,\tau_2)=&~G_{++}(k;\tau_1,\tau_2)-G_{-+}(k;\tau_1,\tau_2),\\
  G_R(k;\tau_1,\tau_2)=&~G_{++}(k;\tau_1,\tau_2)-G_{+-}(k;\tau_1,\tau_2),\\
  F(k;\tau_1,\tau_2)=&-\FR{\ii}{2}\big[G_{+-}(k;\tau_1,\tau_2)+G_{-+}(k;\tau_1,\tau_2)\big].
\end{align}
\eds
But we shall stick to original $\pm$-basis, which is convenient enough for our purpose.

Diagrammatically, we use a black dot and a white dot to denote $+$ point and $-$ point, respectively. Therefore, the four types of propagators can be represented as,
\bgs
\label{prop}
\begin{eqnarray}
\parbox{25mm}{\includegraphics{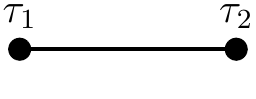}}~&=&~G_{++}(k;\tau_1,\tau_2),\\
\parbox{25mm}{\includegraphics{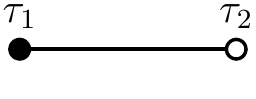}}~&=&~G_{+-}(k;\tau_1,\tau_2),\\
\parbox{25mm}{\includegraphics{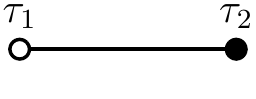}}~&=&~G_{-+}(k;\tau_1,\tau_2),\\
\parbox{25mm}{\includegraphics{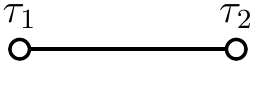}}~&=&~G_{--}(k;\tau_1,\tau_2).
\end{eqnarray}
\eds
The propagators derived above can be applied to both internal legs (or bulk propagators) and external legs (or bulk-to-boundary propagators) of a diagram. For external legs terminated at the final slice $\tau=\tau_f$ (the ``boundary''), we can just take the corresponding argument $\tau$ to $\tau_f$. A boundary point does not distinguish between $+$ and $-$, and thus we have only two types of bulk-to-boundary propagators. We use a square to denote boundary point, then,
\bgs
\label{btobprop}
\begin{eqnarray}
\parbox{25mm}{\includegraphics{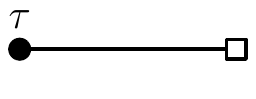}}~&=& G_{+}(k;\tau)\equiv  G_{++}(k;\tau,\tau_f),\\
\parbox{25mm}{\includegraphics{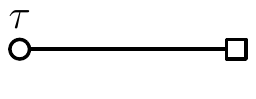}}~&=& G_{-}(k;\tau)\equiv  G_{-+}(k;\tau,\tau_f).
\end{eqnarray}
\eds
In inflation (namely quasi-de Sitter background), the boundary slice $\tau_f$ is the future infinity. Among fields with spin $\leq 2$, the bulk-to-boundary propagators do not vanish only for massless scalar and tensor fields. This is the familiar fact that only nearly massless scalars (including the inflaton) and the helicity-2 graviton survive at the late time limit.

Now we list several frequently used special cases in inflation, i.e. the (Poincar\'e patch of) de Sitter spacetime. We only need to show $G_>$ for each case, as all kinds of propagators can be easily constructed from it.

For a massive scalar field $\si$ of mass $m$ in $(d+1)$-dimensional dS, the mode function is given by,
\begin{align}
  u(\tau;k)=-\ii\FR{\sqrt\pi}{2}e^{\ii\pi(\nu/2+1/4)}H^{(d-1)/2}(-\tau)^{d/2}\text{H}_{\nu}^{(1)}(-k\tau),
\end{align}
where $\nu\equiv \sqrt{d^2/4-(m/H)^2}$. Therefore,
\bge
\label{massiveprop}
\boxed{~~
\begin{aligned}
G_{>}(k;\tau_1,\tau_2)=&~\FR{\pi}{4}e^{-\pi\,\text{Im}\,\nu}H^{d-1}(\tau_1\tau_2)^{d/2}\text{H}_{\nu}^{(1)}(-k\tau_1)\text{H}_{\nu^*}^{(2)}(-k\tau_2).
\end{aligned}
~~}
\ede
When the sound speed $c_s\neq 1$, one just needs to make the substitution $k\to c_s k$.

There are additional two interesting special cases. One is the massless scalar in $(3+1)$-dimensional dS, where we have $m=0$, or equivalently $\nu=3/2$. Then,
\bge
\label{masslessprop}
\boxed{~~
\begin{aligned}
G_{>}(k;\tau_1,\tau_2)=&~\FR{H^2}{2k^3}(1+\ii k\tau_1)(1-\ii k\tau_2)e^{-\ii k(\tau_1-\tau_2)}.
\end{aligned}
~~}
\ede
Note that the propagator for massless scalar does not have the above simple form in general $(d+1)$-dimensions, which makes the dimensional regularization somewhat difficult in this formalism.

The other case is the conformal scalar field in general $(d+1)$-dimensional dS, which has $\nu=1/2$, and thus,
\bge
  G_>(k;\tau_1,\tau_2)=\FR{H^2\tau_1\tau_2}{2k}e^{-\ii k(\tau_1-\tau_2)}.
\ede

\subsubsection{Vertices}

An advantage of staying in the $\pm$ basis for propagators (\ref{SKprop}) rather than going to the Keldysh basis (\ref{KBasis}) is that the diagrammatic rules for interactions are trivially simple. For each single interaction vertex in the original Lagrangian, we only need to write down two vertices, corresponding to $+$ and $-$ type, respectively, and then include an additional minus sign for $-$ type vertex. Several examples should suffice to understand the rules.

\paragraph{Non-derivative couplings.} It is straightforward to write down the rules for non-derivative couplings, just like in ordinary quantum field theory. The only difference is that we Fourier transform the spatial coordinates but not the temporal coordinate, and therefore our diagrammatic rule for vertex looks like a mixed version of coordinate-space Feynman rule and momentum-space Feynman rule. Here we take $\lam\varphi^4$ theory as an example, with the following interaction term,
\bge
  \ld_{\text{int}}\supset-\FR{\lam}{24}a^4(\tau)\varphi^4,
\ede
where $a^4(\tau)$ comes from $\sqrt{-g}$ factor in the Lagrangian. Then we have the following rule for vertices in 3-momentum space,
\begin{align}
&\parbox{22mm}{\includegraphics{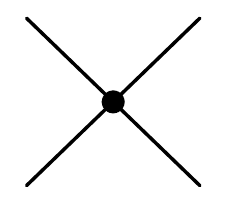}}~=-\ii\lam \int_{\tau_0}^{\tau_f}\di\tau\,a^4(\tau)\cdots,
&\parbox{22mm}{\includegraphics{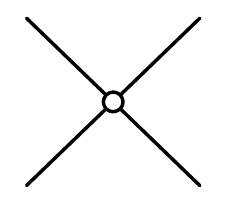}}~=+\ii\lam \int_{\tau_0}^{\tau_f}\di\tau\,a^4(\tau)\cdots,
\end{align}
where $\cdots$ represent all $\tau$-dependent pieces of the diagram coming from the propagators connecting to the vertex.

\paragraph{Derivative couplings.} The vertex containing derivatives of fields can be treated in a similar way. Since we Fourier transform spatial coordinates only, we should consider spatial derivatives and temporal derivatives separately. The spatial derivatives, after Fourier transform, become simple factors of momentum, therefore can be viewed simply as direct coupling, with additional momentum factors. For instance, the diagrammatic rule for the following interaction,
\bge
  \ld_{\text{int}}\supset-\FR{\lam}{6}a^2(\tau)\varphi(\pd_i\varphi)(\pd_i\varphi),
\ede
is given by,
\begin{align}
&\parbox{18mm}{\includegraphics{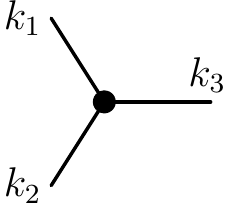}}~~~~~=+\FR{\ii\lam}{3} (\mb k_1\cdot\mb k_2+\mb k_2\cdot\mb k_3+\mb k_3\cdot\mb k_1)\int_{\tau_0}^{\tau_f}\di\tau\,a^2(\tau)\cdots,\\[4mm]
&\parbox{18mm}{\includegraphics{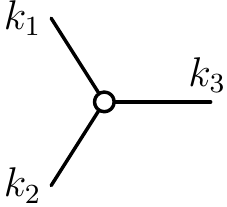}}~~~~~=-\FR{\ii\lam}{3} (\mb k_1\cdot\mb k_2+\mb k_2\cdot\mb k_3+\mb k_3\cdot\mb k_1)\int_{\tau_0}^{\tau_f}\di\tau\,a^2(\tau)\cdots.
\end{align}
On the other hand, the time derivatives appearing in an vertex should be directly applied to the attached propagators. For instance, the rule for the following interaction term,
\bge
  \ld_{\text{int}}\supset-\FR{\lam}{6}a^2(\tau)\varphi\varphi'^2,
\ede
is given by,
\begin{align}
\parbox{18mm}{\includegraphics{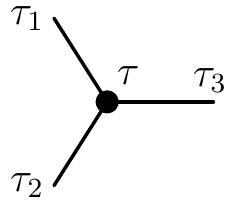}}~~~~~~=&-\FR{\ii\lam}{3}\int_{\tau_0}^{\tau_f}\di\tau\,a^2(\tau)\big[\pd_{\tau}G_{+a_1}(k_1;\tau,\tau_1)\big]\big[\pd_{\tau}G_{+a_2}(k_2;\tau,\tau_2)\big]G_{+a_3}(k_3;\tau,\tau_3)\n\\
      &+\text{2 permutations},
\end{align}
where we have written out the attached three propagators explicitly, and $a_1,a_2,a_3=\pm$. The corresponding rule for minus type vertex is again obtained from the plus type by including an additional minus sign.

The derivative couplings are usually quite subtle in perturbative calculations. In our case the subtlety is that if we apply the time derivative to time-ordered propagator $G_{++}$ or anti-time-ordered propagator $G_{--}$, we must also take account of the time derivative acting on the step function $\theta$ in (\ref{SKprop}), which is crucial for obtaining the correct result. We demonstrate this procedure in a sample calculation of trispectrum in Appendix \ref{sec_SampleDC}.

\subsubsection{Summary and Miscellaneous Discussions}
\label{sec_summary}

With all ingredients for diagrammatic calculations obtained, now we can summarize the diagrammatic rules for calculating expectation values, as follows.

\begin{enumerate}

\item Separate the classical Lagrangian $\ld_\text{cl}[\varphi]$ into free part $\ld_0[\varphi]$ and interaction part $\ld_\text{int}[\varphi]$ as usual. Solve the equation of motion $\de\ld_0[\varphi]/\de\varphi=0$ with initial condition given by $|\Omega\ra$ to find the mode functions of $\varphi$. Due to the spacetime-asymmetric nature of the problem, we Fourier transform the spatial coordinates but not the temporal coordinates. Therefore the mode function can be represented by $u(\tau,\mb k)$, where $\mb k$ is the 3-momentum. Similarly, the external fields $\varphi(\tau,\mb k_i)$ can also be chosen with definite 3-momenta $\mb k_i$.

\item Draw a little square for each $\varphi(\tau,\mb k_i)$ in $Q(\tau)$, which we call external point. Draw interaction vertices as read from $\ld_\text{int}[\phi]$ as usual to desired order in perturbation theory. Then connect all vertices and external points with lines in all possible ways (but no lines between two external points), so that the final diagram is fully connected. Until now, everything is the same as in the ordinary Feynman diagrams.

\item Decorate each vertex by either a black dot (called plus-type vertex) or a white dot (called minus-type vertex), in all possible ways. Therefore we have $2^V$ distinct ways of decorating a diagram with $V$ vertices.

\item Associate the propagator to each line connecting two vertices. Depending on the type of two vertices, there are four types of propagators, as listed in (\ref{prop}). Associate the boundary-to-bulk propagator to each line connecting a vertex and an external line. Depending on the type of bulk vertex, there are two types of boundary-to-bulk propagators, as listed in (\ref{btobprop}). The momentum of each propagator should be chosen such that the total momentum is conserved at each vertex.

\item Associate appropriate factor to each vertex as derived from the Lagrangian, including spatial and temporal derivatives. Add an additional minus sign for each of minus-type vertices. Integrate over all unconstrained and independent 3-momenta. Integrate each vertex over time from initial slice $\tau=\tau_0$ to final slice $\tau=\tau_f$.

\item The symmetric factor is as usual in ordinary Feynman diagrams.

\item The final result for the expectation value $\la Q\ra'$ is the sum of all diagrams, where $\la Q\ra'$ is defined such that $\la Q\ra = (2\pi)^3\delta^3(\sum_i \mathbf{k}_i) \la Q\ra'$.

\end{enumerate}

The diagrammatic rules summarized here will be put in use for quasi-single-field inflation model in next section, which serves as a concrete illustration of the diagrammatic techniques. But at this point, we take a particular expectation value from this model to explain some features of these diagrams. Suppose for now that we have one massless scalar field $\varphi$ and one massive scalar field $\si$, so that $\varphi$ survives in the late time limit. Let us introduce a two-point mixing of the two fields with coupling $\lam_2$, and also a cubic self-interaction for $\si$ with coupling $\lam_3$. To discuss general features of the diagrams, we do not need to specify the explicit form of Lagrangian. Then, if we want to calculate the 3-point function of $\varphi$, at the order of $\lam_2^3\lam_3$, the diagrammatic rules above tell us that we only need to evaluate the following diagram,
\begin{eqnarray}
\label{3ptex}
\parbox{40mm}{\includegraphics{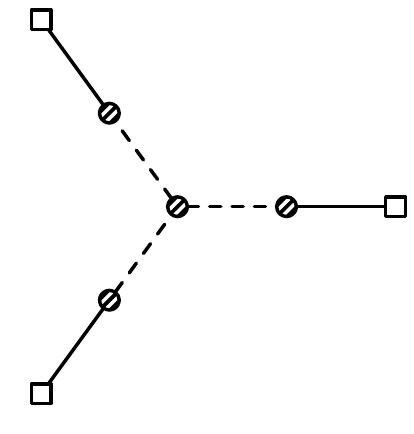}}
\end{eqnarray}
In this diagram, we use solid line and dashed line to represent propagators for $\varphi$ and $\si$, respectively. The external points are marked with little squares, while the shaded dot for each vertex means that it comes with two different types, the plus type (black dot) and the minus type (white dot), and we should sum over all possibilities. This means that the above diagram actually represents the sum of 16 different diagrams since we have 4 shaded dots.

We see that the doubling of fields in the SK formalism, or equivalently, the doubling of vertices in diagrams, significantly complicates the calculation as the number of internal vertices increases. But we do not have to compute all these diagrams one by one, because they are related to each other by complex conjugation. If we go over the diagrammatic rules and the expressions for propagators and vertices, we immediately get the following observation:

\begin{description}
\item[Rule of complex conjugation:] \emph{The complex conjugation of a diagram is obtained by switching black dots to white and vice versa.}
\end{description}

For any expectation value with external points all go to late time limit, which is exactly $\la Q\ra$ considered in this paper, we need to sum over all possible choices of black and white dots. Therefore, an immediate consequence of the above rule of taking complex conjugation is the following:

\begin{description}
\item[Reality of the expectation value:] \emph{$\la Q\ra$ is real.}
\end{description}

For example, the following two diagrams are complex conjugates of each other:
\begin{eqnarray}
\parbox{40mm}{\includegraphics{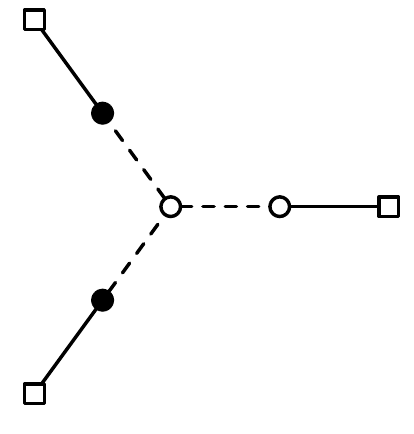}}
   ~~~~~~~~
\parbox{40mm}{\includegraphics{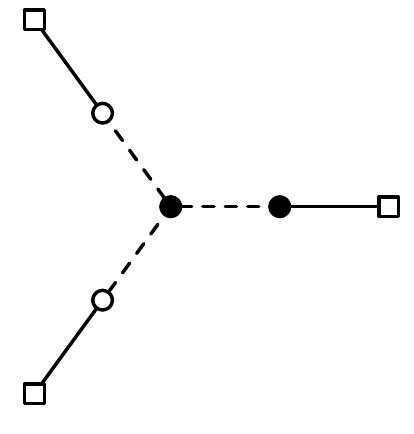}}
\end{eqnarray}
Consequently, for the expectation considered here in (\ref{3ptex}), we only need to calculate 8 diagrams. But this is still quite a lot. Additional trick is needed to further simplify the calculation. The observation here is that the diagram (\ref{3ptex}) contains 3 identical subgraphs with two-point mixing. Therefore, we can firstly evaluate this 2-point mixing subgraph, and then compute the whole diagram. In this way, we reduce the number of diagram from 8 to 1, which turns out to be a significant simplification. This is the idea of ``mixed propagator'' to be introduced in next section. Here we only mention that if the two-point mixing has the form of non-derivative mass mixing,
\bge
  \ld_\text{int}\supset -\lam_2a^{4}(\tau)\varphi\si,
\ede
then it is trivially simple to reduce the 2-point mixing into a single propagator, either by diagonalizing the mass matrix or by direct evaluation using dS covariant techniques like Wick rotation. (See the final appendix of \cite{Chen:2016hrz} for further discussion.) The point of mixed propagator in the next section, however, is to deal with non dS invariant 2-point mixing with derivative coupling, where the dS covariant techniques do not apply, and the diagrammatic rules described here become essential.

\paragraph{Fields with nonzero spin} At the end of this section, we briefly comment on the generalization of the diagrammatic rules to fields with nonzero spins. For bosonic fields, the generalization is straightforward, and the problems such as constraint variables and gauge redundancies can be treated with standard techniques. Suppose we have followed the standard procedure to quantize a spin-$s$ field $\varphi_{\mu_1\cdots\mu_s}$ (with $s$ a positive integer) and have obtained its mode function,
\begin{align}
\varphi_{\mu_1\cdots\mu_s}(\tau,\mb x)=\int\FR{\di^3\mb k}{(2\pi)^3}\sum_{\al}\Big[ u^{\al}(\tau,\mb k)e_{\mu_1\cdots\mu_s}^{\al}(\mb k)b_\al(\mb k) + u^{\al*}(\tau,-\mb k)e_{\mu_1\cdots\mu_s}^{\al*}(-\mb k)b_{\al}^{\dag}(-\mb k)\Big]e^{\ii \mb k\cdot\mb x},
\end{align}
where $\al$ labels helicity states and $e_{\mu_1\cdots\mu_s}^\al(\mb k)$ is the polarization tensor. Then, the  four SK propagators are still given by (\ref{SKprop}), with
\begin{align}
(G_>)_{\mu_1\cdots\mu_s,\nu_1\cdots\nu_s}=\sum_{\al}u_{\al}(\tau_1,\mb k)u_{\al}^*(\tau_2,\mb k)e_{\mu_1\cdots\mu_s}^{\al}(\mb k)e_{\nu_1\cdots\nu_s}^{\al*}(\mb k).
\end{align}
A particularly useful example is the massless graviton, which appears as the tensor mode in primordial perturbations. The two independent helicity states of the graviton can be described by the transverse and traceless part of a spatial rank-2 tensor $\ga_{ij}$, with the following Lagrangian at leading order,
\bge
  \ld_\text{cl}\supset \FR{\Mp^2}{8}a^2(\tau)\Big[(\pd_\tau\ga^i{}_j)(\pd_\tau\ga^j{}_i)-(\pd_k\ga^i{}_j)(\pd_k\ga^j{}_i)\Big].
\ede
The corresponding propagator is given by,
\begin{align}
  (G_>)_{ij,k\ell}=&~\FR{2}{\Mp^2}\sum_{\al}u_\al(\tau_1,\mb k)u_\al^*(\tau_2,\mb k)e^{\al}_{ij}(\mb k)e_{k\ell}^{\al*}(\mb k)\n\\
  =&~\FR{H^2}{\Mp^2k^3}(1+\ii k\tau_1)(1-\ii k\tau_2)e^{-\ii k(\tau_1-\tau_2)}\sum_{\al}e_{ij}^\al(\mb k)e_{k\ell}^{\al*}(\mb k).
\end{align}
More details on mode functions of bosonic higher spin fields during inflation can be found in \cite{Lee:2016vti}.

For fermionic fields, one can still use the formalism described in this section to develop a set of diagrammatic rules. The complication here is that the anti-commuting nature of fermions forbids us to mix the path integral over plus fields and minus fields. As a result, one cannot use formulae such as (\ref{QevHPI}) for fermions. More discussion about how to quantize fermions properly in this formalism can be found in \cite{Jordan:1986ug} and also in \cite{Chen:2016nrs}.

\section{Application to Quasi-Single-Field Inflation}
\label{sec_QSFI}

In this section, we use quasi-single-field inflation \cite{Chen:2009we,Chen:2009zp} as an example to show how the diagrammatic method leads to neat results for 3-point and 4-point correlation functions of scalar perturbation.

The quasi-single-field inflation in general refers to the inflation scenarios with one or more spectator fields of mass around Hubble scale. As a simple example, we consider the model with a slightly curved inflation trajectory described by the following action with two real scalar fields $\theta$ and $\si$,
\bge
  S=\int\di^4x\sqrt{-g}\bigg[-\FR{1}{2}(\wt{R}+\si)^2(\pd_\mu\theta)^2-\FR{1}{2}(\pd_\mu\si)^2-V_{\text{sr}}(\theta)-V(\si)\bigg],
\ede
where $V_\text{sr}(\theta)$ is an arbitrary slow-roll potential, while $V(\si)$ is a potential such that $\si$ obtains a classical constant background $\si_0$. After expanding the fields around their classical background $\theta_0$ and $\si_0$, the Lagrangian for the fluctuation field has the following form,
\begin{align}
\ld_\text{cl}=&\,\FR{a^2}{2}\Big((\de\phi')^2-(\pd_i\de\phi)^2+(\de\si')^2-(\pd_i\de\si)^2\Big)-\FR{a^4m^2}{2}\de\si^2\n\\
&~+a^3\lam_2\de\si\de\phi'-a^4\Big(\FR{\lam_3}{6}\de\si^3+\FR{\lam_4}{24}\de\si^4+\cdots\Big),
\end{align}
where we have defined $\de\phi=(\wt R+\si_0)\de\theta$, and the scale factor $a(\tau)\simeq -1/(H\tau)$. The first line of the above Lagrangian can be identified as free part $\ld_0$, with a massless scalar $\de\phi$ and a massive scalar $\de\si$ of mass $m^2=V''(\si_0)-\dot\theta_0^2$. In the second line, we have interactions with two-point derivative mixing between $\de\phi$ and $\de\si$, with coupling strength $\lam_2=2\dot\theta_0$, as well as self-interactions of $\de\si$, with couplings $\lam_3=V'''(\si_0)$ and $\lam_4=V^{(4)}(\si_0)$. The self-interactions of $\de\si$ are not constrained by slow-roll conditions, and thus can be large.

Below we shall compute the leading order correction to the power spectrum, as well as the leading bispectrum and trispectrum, using our diagrammatic method. In these computations, the two-point mixing vertex appears frequently. Therefore, it turns out to be a great simplification if we firstly isolate the two-point mixing and evaluate it into a closed form, which we call a ``mixed propagator''. Therefore, before computing various correlation functions, we devote the next subsection to the evaluation of the mixed propagator.

\subsection{Mixed Propagator}

By mixed propagator, we mean the following object,
\bge
\parbox{85mm}{\includegraphics{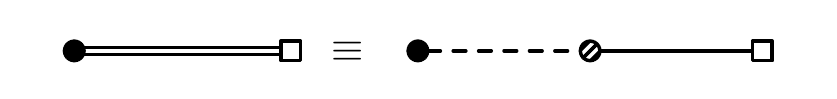}}
\ede
That is, we consider a two-point function $\la\de\si\de\phi\ra$, with $\de\phi$ going to late time limit. This is a bulk-to-boundary propagator, with only one endpoint depending on time. For clarity, we denote the propagator of $\de\phi$ by $G(k;\tau_1,\tau_2)$ and the propagator of $\de\si$ by $D(k;\tau_1,\tau_2)$. Then, using the diagrammatic rules, in particular (\ref{massiveprop}) and (\ref{masslessprop}), the above mixed propagator can be translated into the following expression,
\begin{align}
\mathcal{G}_\pm(k;\tau)=&~\ii\lam_2 \int_{-\infty}^0 \FR{\di\tau'}{(-H\tau')^3}\Big[D_{\pm+}(k;\tau,\tau')\pd_{\tau'}G_{+}(k;\tau')-D_{\pm-}(k;\tau,\tau')\pd_{\tau'}G_{-}(k;\tau')\Big]\n\\
=&~\FR{\pi\lam_2 H}{8k^3}I_\pm(z),
\end{align}
where $z\equiv -k\tau$, and $I_\pm(z)$ is given by,
\begin{align}
\label{Ipm}
I_\pm(z)=&~e^{-\pi\,\text{Im}\,\nu}z^{3/2}\bigg\{2\,\text{Im}\bigg[\mathrm{H}_\nu^{(1)}(z)\int_0^{\infty}\FR{\di z'}{\sqrt{z'}} \mathrm{H}_{\nu^*}^{(2)}(z')e^{-\ii a z'}\bigg]\n\\
&~+\ii \mathrm{H}_\nu^{(1)}(z)\int_0^z\FR{\di z'}{\sqrt{z'}} \mathrm{H}_{\nu^*}^{(2)}(z')e^{\mp\ii  z'}-\ii \mathrm{H}_{\nu^*}^{(2)}(z)\int_0^z\FR{\di z'}{\sqrt{z'}} \mathrm{H}_\nu^{(1)}(z')e^{\mp\ii  z'}\bigg\},
\end{align}
where in the first line we have inserted a parameter $a=1-\ii\ep$ with real part 1 and a small negative imaginary part to take care of $\ii\ep$ prescription. The integral can be carried out explicitly, as follows,
\begin{align}
I_\pm(z)
=&~z^{3/2}e^{-\pi\,\text{Im}\,\nu}\bigg\{\Big[\mathcal{C}_\nu+(\cot(\pi\nu)-\ii)f_\nu^\pm(z)-\csc(\pi\nu)f^\pm_{-\nu}(z)\Big]\mathrm{H}_{\nu^*}^{(2)}(z)\n\\
&~+\Big[\mathcal{C}_\nu^*+(\cot(\pi\nu^*)+\ii)f^\pm_{\nu^*}(z)-\csc(\pi\nu^*)f^\pm_{-\nu^*}(z)\Big]\mathrm{H}_{\nu}^{(1)}(z)
\bigg\},
\end{align}
where $f^\pm_\nu(z)$ is defined by,
\begin{align}
\label{fnu}
  f^\pm_\nu (z)=\FR{z^{\nu+1/2}}{2^\nu(\nu+1/2)\Gamma(\nu+1)}{}_2F_2\Big(\nu+\FR{1}{2},\nu+\FR{1}{2};\nu+\FR{3}{2},2\nu+1;\mp 2\ii z\Big),
\end{align}
and $\mathcal{C}_\nu$ is a $z$-independent coefficient, given by,
\begin{align}
\label{cnu}
\mathcal{C}_\nu=&~\ii\int_0^\infty\FR{\di z}{\sqrt{z}}H_\nu^{(1)}(z)e^{+\ii a z}=\sqrt{2\pi}e^{\ii\pi(1/4-\nu/2)}\sec(\pi\nu).
\end{align}
where we have taken $a\to 1$ limit in the final result, and thus the UV convergence is manifest.

\subsection{Leading Correction to Power Spectrum}

As a first application of mixed propagator, we calculate the leading order correction to the power spectrum from 2-point mixing. Using the mixed propagator, we need to calculate the following diagram,
\bge
\parbox{85mm}{\includegraphics{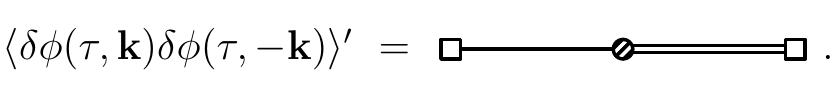}}
\ede
Using the diagrammatic rules, we can write down the corresponding expression immediately,
\begin{align}
&\la\de\phi(\tau,\mb k)\de\phi(\tau,-\mb k)\ra'\n\\
=&~\ii\lam_2 \int_{-\infty}^0\FR{\di\tau'}{(-H\tau')^3}\Big[\pd_{\tau'}G_{+}(k;\tau')\mathcal{G}_+(k;\tau')-\pd_{\tau'}G_{-}(k;\tau')\mathcal{G}_-(k;\tau')\Big]\n\\
=&~\FR{\lam_2^2}{k^3}\mathcal{P}(\nu),
\end{align}
where $\mathcal{P}(\nu)$ is the following integral and can be carried out completely as was done in \cite{Chen:2012ge},
\begin{align}
  \mathcal{P}(\nu)
\equiv \FR{-\ii \pi}{16}\int_0^\infty\FR{\di z}{z^2}\Big[e^{-\ii z}I_+(z)-e^{+\ii z}I_{-}(z)\Big]
  = \FR{\pi^2}{4\cos^2(\pi\nu)}+\Theta(\nu)+\Theta(-\nu),
\end{align}
where the function $\Theta(\nu)$ is defined to be,
\bge
  \Theta(\nu)\equiv
  \text{Im} \left\{
  \FR{e^{-\ii\pi\nu}}{16\sin(\pi\nu)}\bigg[\psi^{(1)}\Big(\FR{1}{4}+\FR{\nu}{2}\Big)-\psi^{(1)}\Big(\FR{3}{4}+\FR{\nu}{2}\Big)\bigg]\right\},
\ede
where $\psi^{(1)}(z)\equiv\di^2\log\Gamma(z)/\di z^2$. The result obtained here using the diagrammatic rules agrees trivially with the previous calculation in canonical in-in formalism \cite{Chen:2012ge}, as it should. The diagrammatic method is not more advantageous in this calculation as the quantity itself is simple enough. But it is still helpful to use this warm-up exercise as a check of the diagrammatic method.

\subsection{Bispectrum}

\begin{figure}[tbph]
\centering
\includegraphics[width=0.3\textwidth]{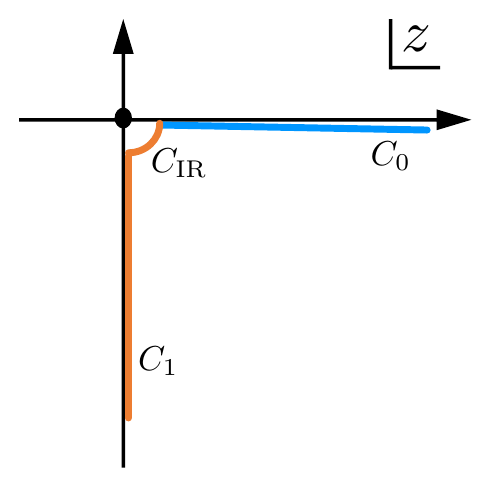}
\hspace{1cm}
\includegraphics[width=0.3\textwidth]{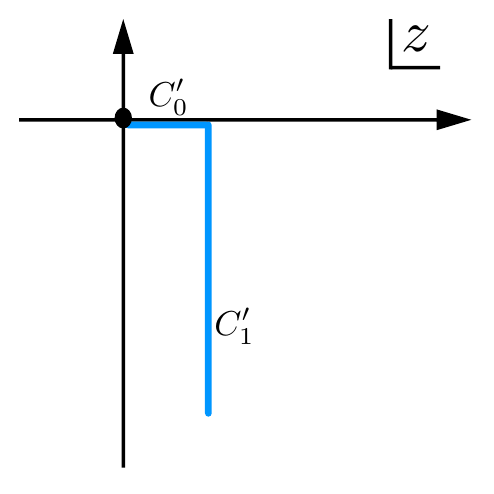}
\caption{Integral contour in (\ref{3pt}). The left panel is helpful for a theoretical understanding of IR divergence, while the right panel is suitable for numerical calculation.}
\label{fig_contour}
\end{figure}

The real power of mixed propagator can only be appreciated when we compute non-Gaussianities, i.e. the bispectrum and even the trispectrum. The leading contribution to the 3-point function is from the contribution of order $\lam_2^3\lam_3^{}$. We have met this expectation value (\ref{3ptex}) in last section. Now using mixed propagator, we can recast it into the following form,
\begin{eqnarray}
\parbox{40mm}{\includegraphics{FD_3pt_1}}~~~~=~~
\parbox{40mm}{\includegraphics{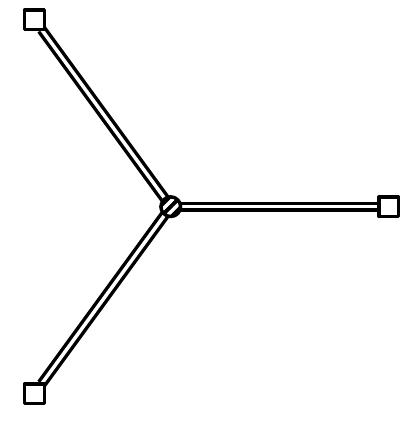}}\end{eqnarray}
In this way, we have reduced the number of internal vertices from 4 to 1. Therefore, rather than summing over $2^4=16$ diagrams, we only need to consider $2$ diagrams. In addition, these two diagrams are complex conjugates of each other, and thus we can immediately write down the following expression for this diagram,
\begin{align}
\label{3pt}
\la\de\phi(\tau,\mb k_1)\de\phi(\tau,\mb k_2)\de\phi(\tau,\mb k_3)\ra_{\lam_3}'=&~2\lam_3\,\text{Im}\int_{-\infty}^0 \FR{\di\tau}{(-H\tau)^4}\mathcal{G}_+(k_1;\tau)\mathcal{G}_+(k_2;\tau)\mathcal{G}_+(k_3;\tau)\n\\
=&~\FR{\pi^3\lam_2^3\lam_3^{}}{256Hk_2^3k_3^3}\,\text{Im}\int_0^\infty\FR{\di z}{z^4}I_+(z)I_+(\FR{k_2}{k_1}z)I_+(\FR{k_3}{k_1}z).
\end{align}
This time it is very challenging, if not completely impossible, to carry out the integral analytically. But the integral can be readily done numerically, and the numerical computation for this quantity is much faster than doing the integral obtained from canonical in-in formalism, because the four-layer integrals in \cite{Chen:2009we,Chen:2009zp} are neatly organized and reduced into a one-layer integral due to the use of the mixed propagator. The only subtlety for numerical calculation is the implementation of $\ii\ep$-prescription. Ideally, one may do Wick rotation to move the original slightly deformed contour along the real axis ($C_0$ in the left panel of Fig.\;\ref{fig_contour}) entirely to the imaginary axis ($C_1$ in the same plot), so that the UV oscillations are exponentially damped. However, in the case of (\ref{3pt}), we have the problem of IR divergence, because the integrand is singular as $z\to 0$ when $\nu\geq 1/2$. This singularity exists only in the real part of the integrand, and thus should disappear when we take the imaginary part in (\ref{3pt}). However, the situation gets changed if we do the Wick rotation. In fact, to avoid the singularity, one must carefully go around $z=0$ by including a small arc ($C_\text{IR}$ in the same plot) when doing Wick rotation. We can easily check that the integral along this small arc is actually divergent as its radius shrinks to zero, when $\nu>1/2$. In fact, as $z\to 0$, the function $I_+(z)$ defined in (\ref{Ipm}) behaves like $I(z)\sim z^{3/2-\nu}$, and therefore the integrand of (\ref{3pt}) goes like $z^{1/2-3\nu}$. Thus the integral along the small arc behaves like $r^{3/2-3\nu}$, where $r$ is the radius of the arc. Such divergent IR behavior is rather difficult to handle numerically. Therefore a more practical way of numerical evaluation is to avoid IR divergent 4th quadrant close to $z=0$, and choose a contour like $C_0'+C_1'$ as shown on the right panel of Fig.\;\ref{fig_contour}. In this way, both UV oscillation and IR divergence are avoided properly.

This contour is similar to the ``mixed form integration" and ``shifted Wick rotation" used in Ref.~\cite{Chen:2009zp} (see Appendix B.3 and Appendix C of \cite{Chen:2009zp} for explanation), but the integral here is organized in a much more economic form.
In the canonical approach, due to the explicit time-ordered integrals, it is much harder to separate the mixed propagator from the rest of the in-in integral (unless rewriting the in-in integral into a SK-like form). Thus the SK formalism makes it much easier to reuse common sub-diagrams, for simplification and resummation purposes.

\begin{figure}[tbph]
  \centering
  \includegraphics[width=0.75\textwidth]{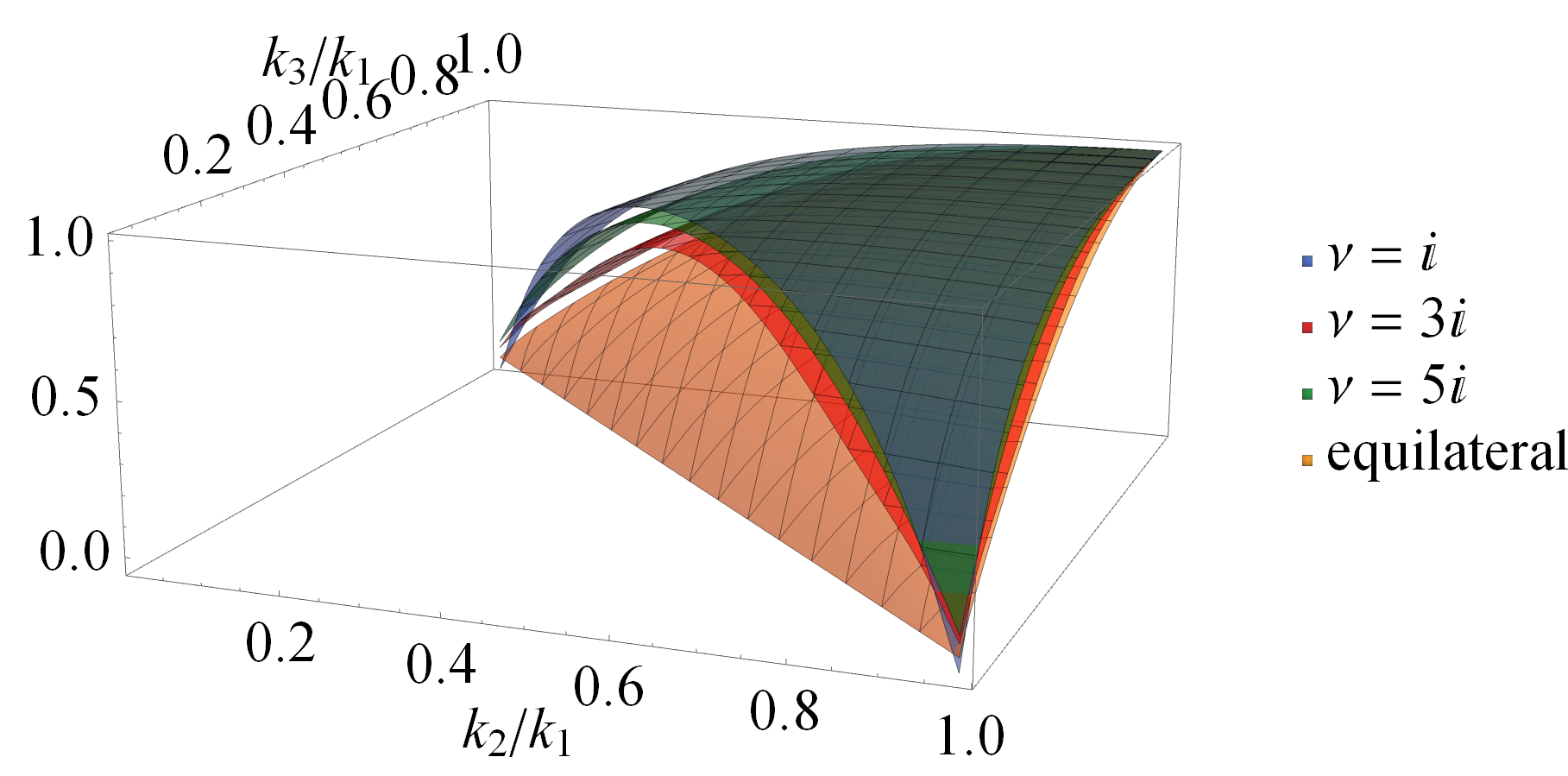}
  \caption{\label{fig:im3d} We plot the dimensionless shape functions. Near the equilateral limit $k_1=k_2=k_3$, from top to down, the top 3 layers are $\nu=i$ (blue), $\nu=5i$ (green), $\nu=3i$ (red), respectively. Here the shape functions are normalized to be 1 at the equilateral limit. In the equilateral limit, before normalization, the three-point function is not monotonic as a function of $i\nu$. Thus the layers with different values of $i\mu$ intersects with each other due to this normalization. We also plotted the factorizable ansatz of the equilateral shape \cite{Creminelli:2005hu} (orange). This ansatz is a good approximation of, but not identical to, the equilateral shape originating from models with a small sound speed. The equilateral shape (from small sound speed models) would look more similar to the other layers of the plot.}
\end{figure}

\begin{figure}[tbph]
  \centering
  \includegraphics[width=0.85\textwidth]{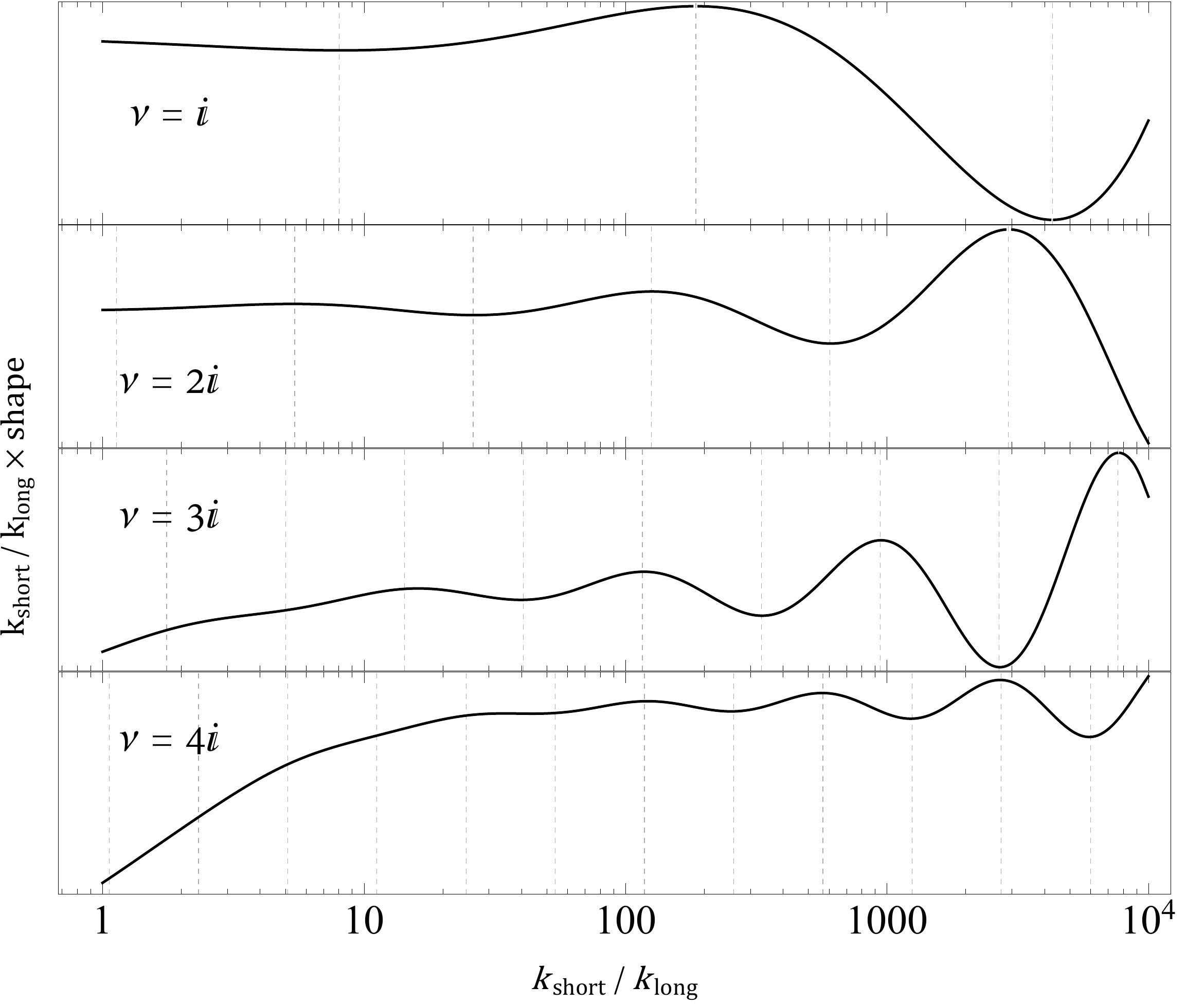}
  \caption{\label{fig:sqi} The squeezed limits for $\nu=i,2i,3i,4i$. The vertical axis is the dimensionless shape function $S$ magnified by a factor of $k_\mathrm{short}/k_\mathrm{long}$ for visual effect. The oscillatory components, namely the clock signals \cite{Chen:2015lza}, take the same form as those in Fig.4 of Ref.~\cite{Chen:2015lza} but their relative amplitudes to the non-oscillatory components are larger. This is because the couplings used in these two examples are different. In \cite{Chen:2015lza}, a simple example of coupling is studied; while here we have used the full leading coupling of the quasi-single-field model in \cite{Chen:2009we,Chen:2009zp}.}
\end{figure}

The form of \eqref{3pt} is very convenient for numerical calculation of the dimensionless shape function $S(k_1,k_2,k_3)$, defined as,
\bge
\label{shape}
  \la\zeta(\mb k_1)\zeta(\mb k_2)\zeta(\mb k_3)\ra'\equiv (2\pi)^4S(k_1,k_2,k_3)\FR{1}{(k_1k_2k_3)^2}P_\zeta^2,
\ede
where $\zeta$ is the curvature perturbation, which is related to the inflaton fluctuation via $\zeta =-H\de\phi/\dot\phi_0$, and $P_\zeta=H^2/(8\pi^2\Mp^2\ep)$ is the power spectrum of $\zeta$.
The shape functions $S$ for some imaginary values of $\nu$ are plotted in Fig.~\ref{fig:im3d}, and some squeezed limit examples are plotted in Fig.~\ref{fig:sqi}. (The case of real $\nu$ can be found at \cite{Chen:2009zp}.%
\footnote{For real $\nu$, the leading contribution in the squeezed limit is
\begin{align}
  S(k_1,k_2,k_3)= P_\zeta^{-1/2}
  \left(\frac{\lambda_2}{H}\right)^3 \left ( \frac{\lambda_3}{H} \right ) \left [ -\frac{1}{12} s(\nu) \right ] \left ( \frac{k_3}{k_1} \right )^{1/2-\nu},
\end{align}
where we have normalized $s(\nu)$ such that it agrees with the definition in Eq.~(5.10) of \cite{Chen:2009zp} , and in our context $s(\nu)$ takes the form
\begin{align}
  s(\nu) = - \frac{3\times 2^{-6+\nu}\pi^{5/2}\Gamma(\nu)}{\cos(\pi\nu/2)+\sin(\pi\nu/2)}
  \Im\left[ \int_0^\infty\di z\,I_+^2(z)z^{-5/2-\nu} \right]~.
\end{align}
This is written in a much more economic form than \cite{Chen:2009zp}.
})

 The squeezed limit of bispectrum may be separated into an equilateral component and an oscillatory/power-law component. These are contributed by the local and non-local component of the massive field propagator, respectively \cite{Arkani-Hamed:2015bza}. The ``local" component of the propagator is analytic in momentum and vanishes as the momentum goes to zero. The ``non-local" component is non-analytic in momentum and gives rise to long range correlation in position space. (Note that this terminology is used in a different context from the terminology ``local" used in describing the local bispectrum.)
The oscillatory component of the bispectrum has a special physical significance \cite{Chen:2015lza}. These signals are essentially the imprints of the standard oscillation of massive fields (the ``primordial standard clock") and are called the ``clock signals" . They directly measure the scale factor of the primordial universe as a function of time $a(t)$ and can be used as a direct evidence to distinguish the inflation scenario from the alternatives.

\begin{figure}[tbph]
  \centering
  \includegraphics[width=0.48\textwidth]{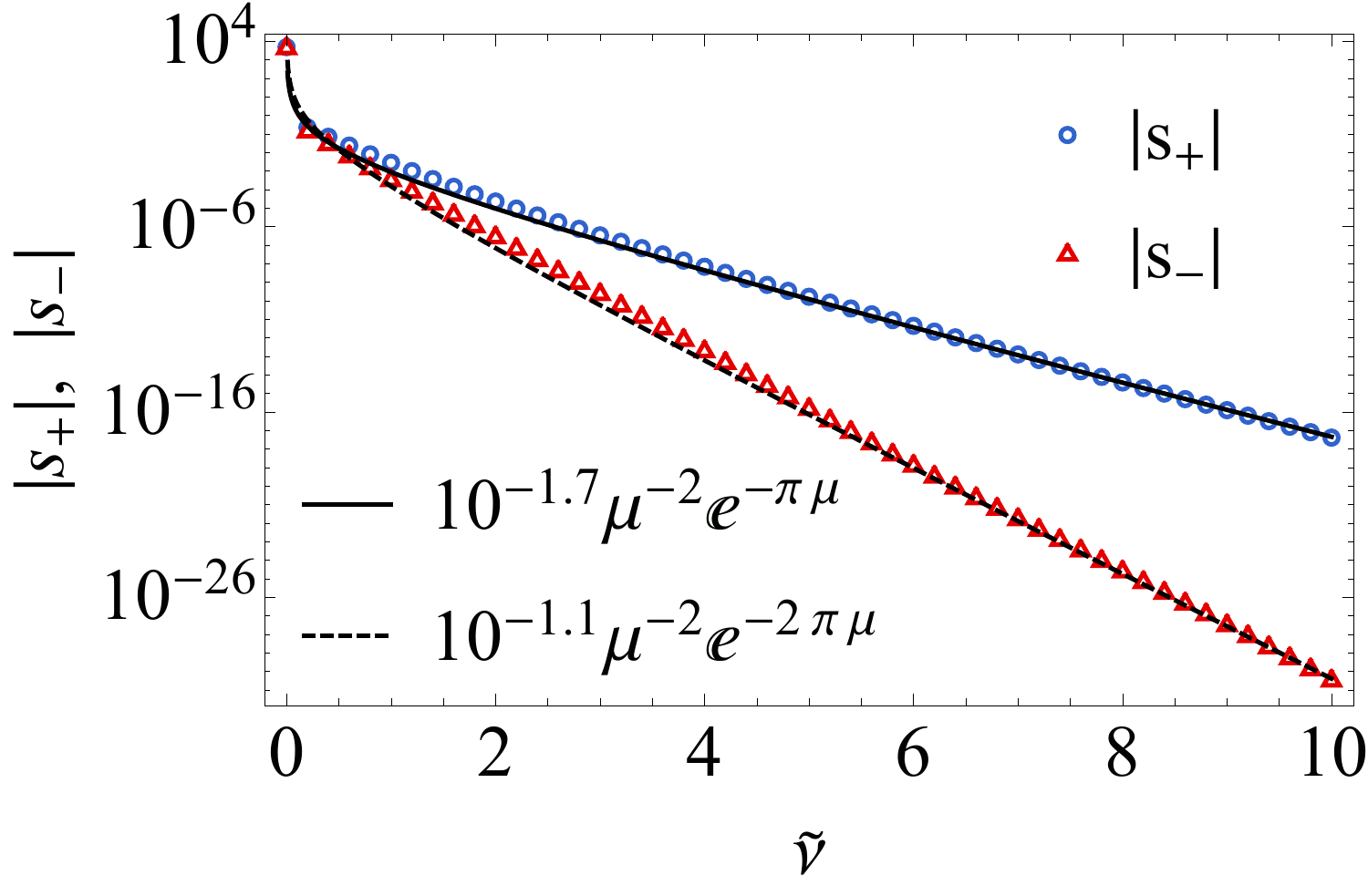}
  \includegraphics[width=0.48\textwidth]{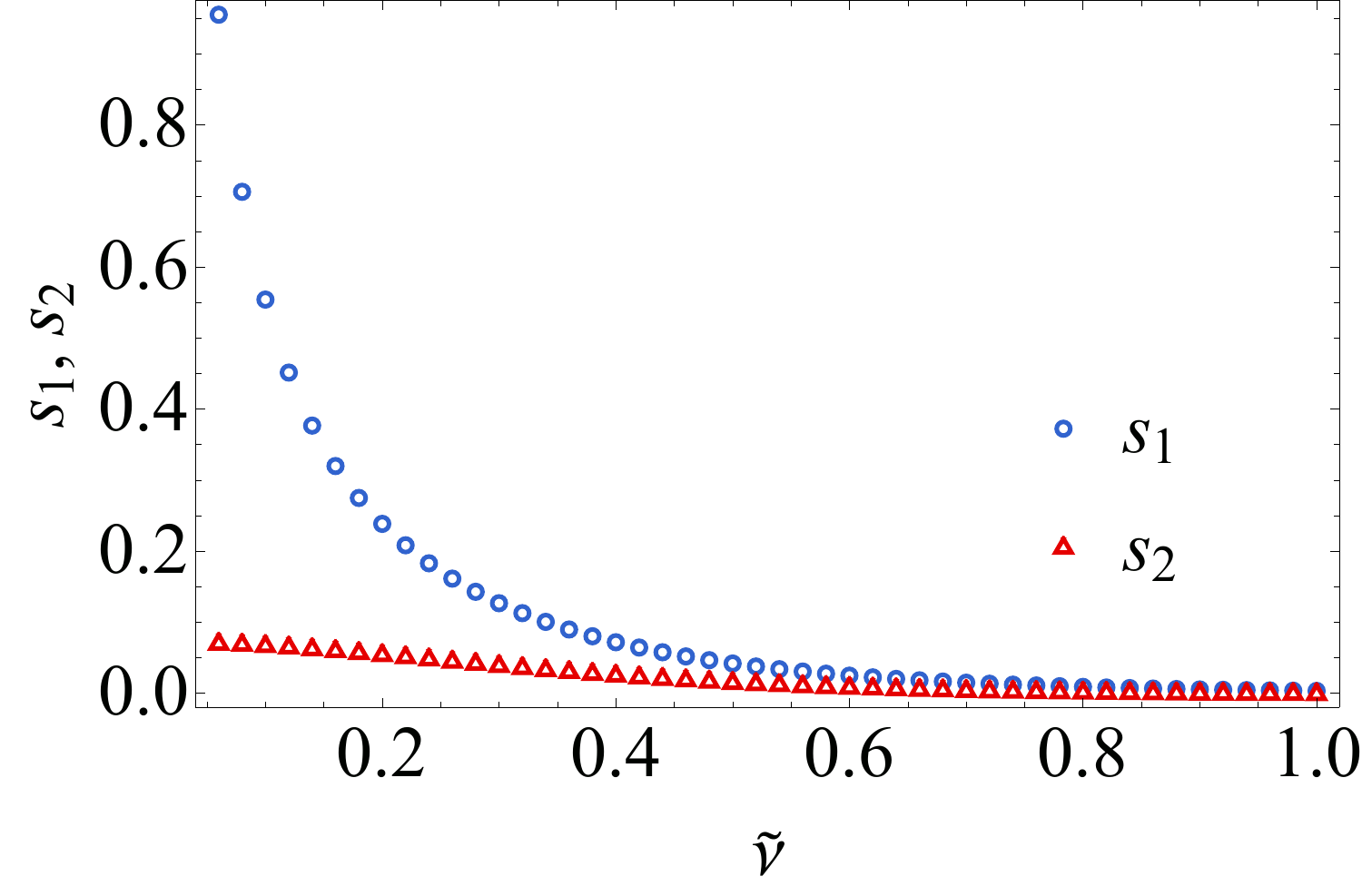}
  \includegraphics[width=0.48\textwidth]{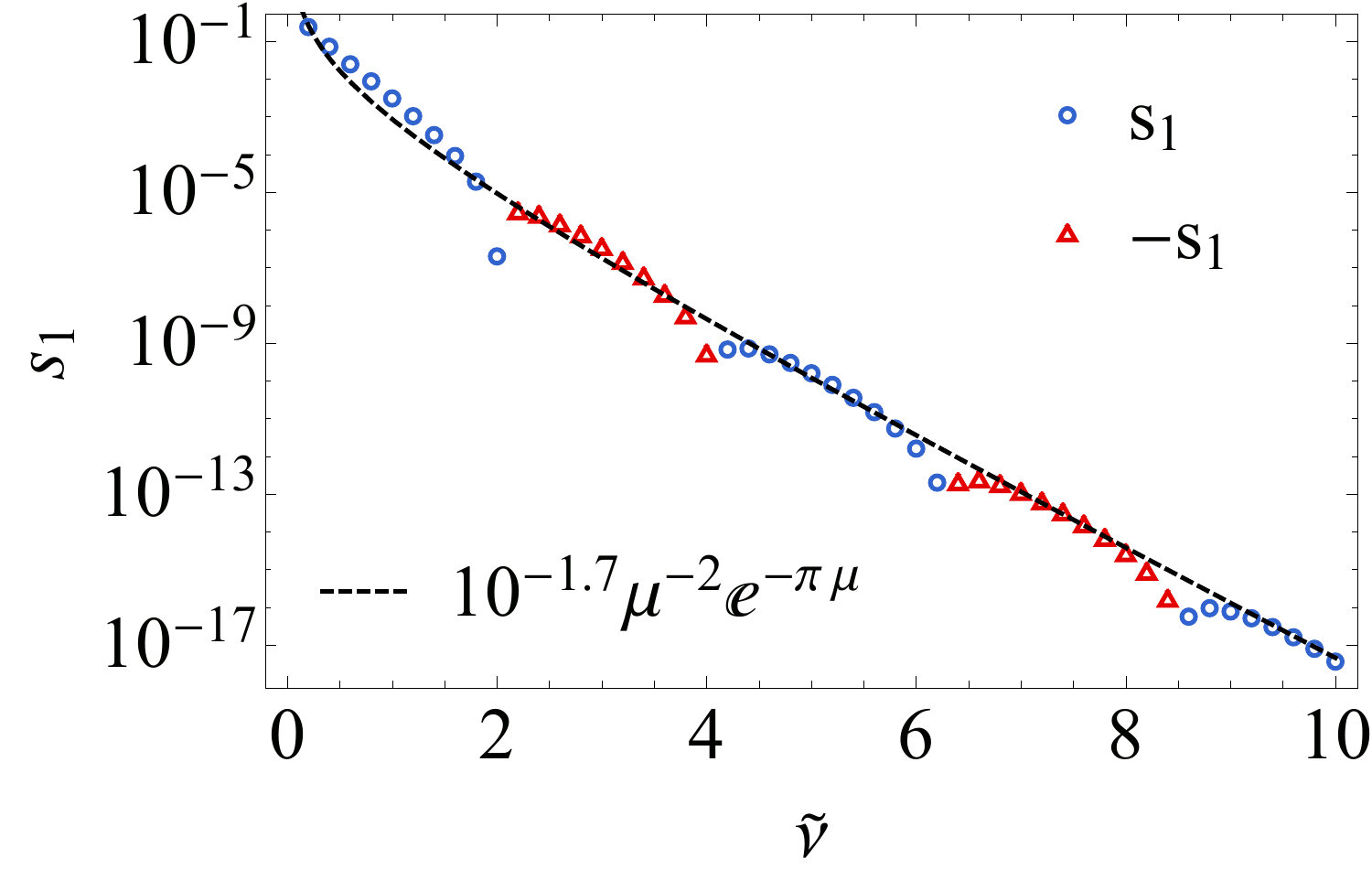}
  \includegraphics[width=0.48\textwidth]{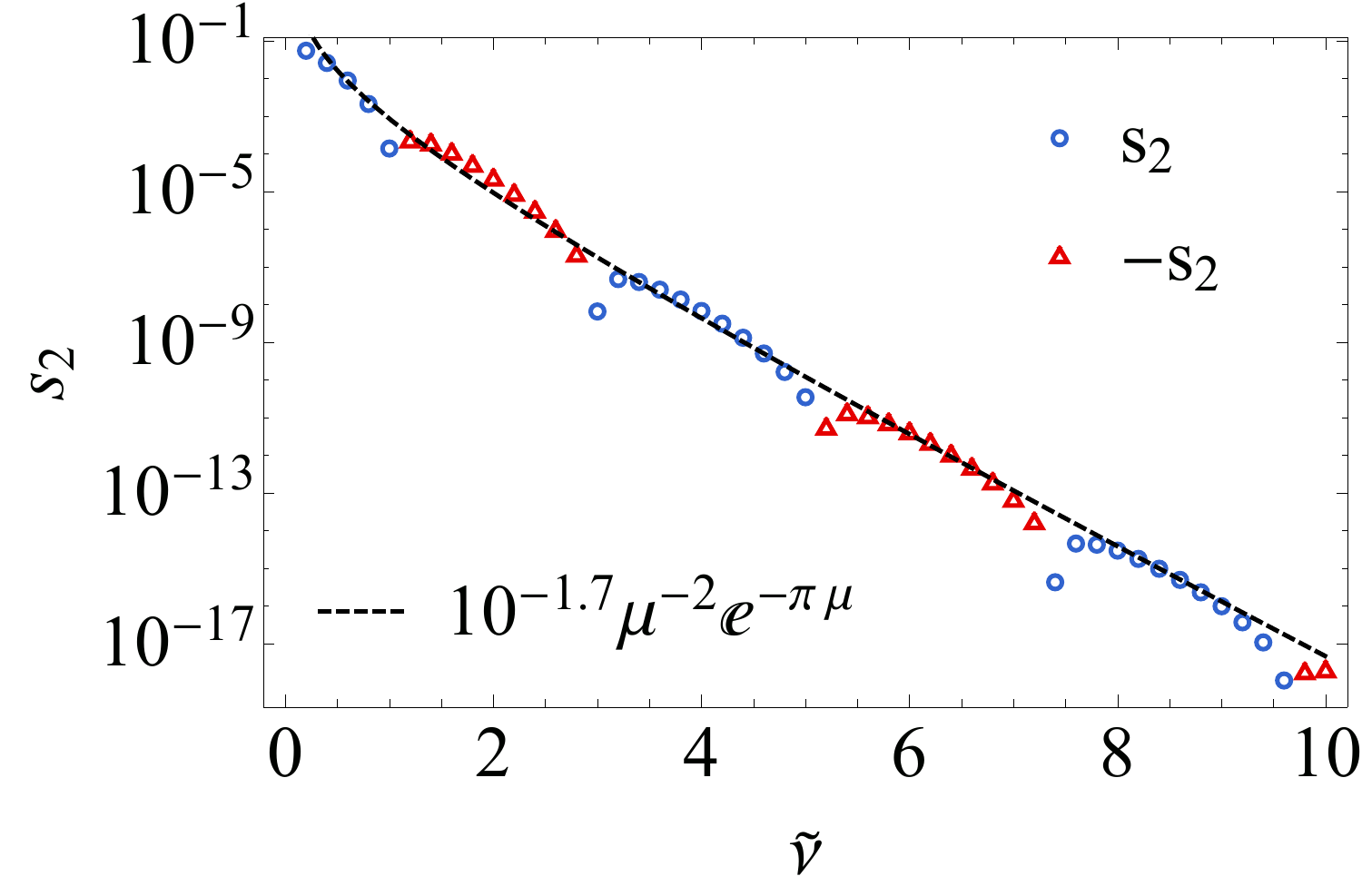}
  \caption{\label{fig:sqabs}
  \textbf{Upper-left panel:} $s_\pm(\wt\nu)$. Numerically, one finds that at $\wt\nu \gg 1$, $s_+(\wt\nu) \sim e^{-\pi\wt\nu}$, and $s_-(\wt\nu) \sim e^{-2\pi\wt\nu}$ up to polynomial factors (the power of the polynomial is chosen to better fit the curve overall for the range of $\wt\nu$ plotted). As noticed in \cite{Chen:2009zp}, the apparent divergence at $\wt\nu \rightarrow 0$ is an indication of change of shape, instead of anything physical blowing up.
  \textbf{Upper-right panel:} $s_{\{1,2\}}(\wt\nu)$ for $0<\wt\nu<1$.
  \textbf{Lower-left panel:} Large $\wt\nu$ behavior for $s_1(\wt\nu) $.
  \textbf{Lower-right panel:} Large $\wt\nu$ behavior for $s_2(\wt\nu) $.}
\end{figure}

To isolate the oscillatory component, namely the clock signal, from the bispectrum, we expand (\ref{3pt}) in $k_3/k_1\to 0$ limit,%
\footnote{More precisely, in order for the oscillatory component to dominate over the non-oscillatory one, we need $k_3/k_1 \ll e^{-2\pi\wt\nu}$. Also, the condition $\wt\nu$ is real is used in this expansion.}
and use (\ref{shape}) to write the shape function as follows,
\begin{align}
  S(k_1,k_2,k_3) \to P_\zeta^{-1/2}
  \left(\frac{\lambda_2}{H}\right)^3 \left ( \frac{\lambda_3}{H} \right )
  \text{Im}\,\bigg[
    s_+(\wt\nu)\Big(\FR{k_3}{k_1}\Big)^{1/2+\ii\wt\nu}
  + s_-(\wt\nu)\Big(\FR{k_3}{k_1}\Big)^{1/2-\ii\wt\nu}\bigg],
\end{align}
where we have defined $\nu=\ii\wt\nu$ so that $\wt\nu$ is real for $m>3H/2$, and the coefficients $s_\pm(\wt\nu)$ are given by,
\begin{align}
  s_+(\wt\nu)=&~\FR{-2^{-\ii\wt\nu}\pi^{5/2}}{256\Gamma(1+\ii\wt\nu)\sinh(\pi\wt\nu)\big[\sinh(\pi\wt\nu/2)+\ii\cosh(\pi\wt\nu/2)\big]}\int_0^\infty\di z\,I_+^2(z)z^{-5/2+\ii\wt\nu},\\
  s_-(\wt\nu)=&~\FR{-2^{+\ii\wt\nu}\pi^{5/2}}{256\Gamma(1-\ii\wt\nu)\sinh(\pi\wt\nu)\big[\sinh(\pi\wt\nu/2)-\ii\cosh(\pi\wt\nu/2)\big]}\int_0^\infty\di z\,I_+^2(z)z^{-5/2-\ii\wt\nu}.
\end{align}
The functions $s_\pm(\wt\nu)$ are plotted in the upper-left panel of Fig.~\ref{fig:sqabs}. At large $\wt\nu$, $s_+(\wt\nu) \sim e^{-\pi\wt\nu}$, which is the dominate contribution in the oscillatory component in the squeezed limit. This agrees with the observation in \cite{Arkani-Hamed:2015bza, Chen:2015lza}, where a simplified model is studied\footnote{Here we have three massive propagators, but for the leading contribution to the squeezed limit oscillations, only one massive propagator is chosen to be non-local and thus exponentially suppressed. The other two massive propagators are local and thus contribute power-law suppression factors instead of exponential. The terms with more ``non-local'' propagators are suppressed by $e^{-2\pi\wt\nu}$ and $e^{-3\pi\wt\nu}$ and thus are subdominant. Note that if we take all three propagators to be ``local'', then one gets power-law suppression instead of exponential \cite{Gong:2013sma}, with equilateral shape of non-Gaussianity. This equilateral component of non-Gaussianity cannot be distinguished from single field inflation with a modified sound speed \cite{Chen:2006nt} (and its amplitude decreases faster in the squeezed limit than the oscillatory signal that we are discussing). Thus the oscillatory component of the non-Gaussianity, with its amplitude scales as $e^{-\pi\wt\nu}$, is the leading distinctive signature of canonical massive scalar fields with $m>3H/2$ during inflation.}.

To better explore the properties of $S(k_1,k_2,k_3)$ in the squeezed limit, it is convenient to write
\begin{align}
  S(k_1,k_2,k_3)= P_\zeta^{-1/2}
  \left(\frac{\lambda_2}{H}\right)^3 \left ( \frac{\lambda_3}{H} \right ) \Big(\FR{k_3}{k_1}\Big)^{1/2}
  \left \{
      s_1(\wt\nu) \sin\left [ \wt\nu \log\left ( \frac{k_3}{k_1}  \right ) \right ]
    + s_2(\wt\nu) \cos\left [ \wt\nu \log\left ( \frac{k_3}{k_1}  \right ) \right ]
  \right \},
\end{align}
where
\begin{align}
  s_1(\wt\nu) = \Re (s_+ - s_-),\qquad
  s_2(\wt\nu) = \Im (s_+ + s_-).
\end{align}
The behavior of $s_1(\wt\nu)$ and $s_2(\wt\nu)$ are plotted in Fig.~\ref{fig:sqabs} on the upper-right panel for $0<\wt\nu<1$, and the lower panels for larger $\wt\nu$.

Because $P_\zeta^{-1/2} \sim 10^5$, the amplitude of the clock signal in this bispectrum can be observably large \cite{Meerburg:2016zdz}. Thus we have presented a concrete example of ``quantum primordial standard clock" models \cite{Chen:2015lza}, in which the clock signal can be large and potentially detectable.

\subsection{Trispectrum}

The same method of calculating bispectrum can be well applied to the case of trispectrum. In this case, we have two comparable types of contributions. One is from the quartic self-interaction of $\si$ field, at order $\lam_2^4\lam_4^{}$, which is described by the following diagram,
\bge
\parbox{105mm}{\includegraphics{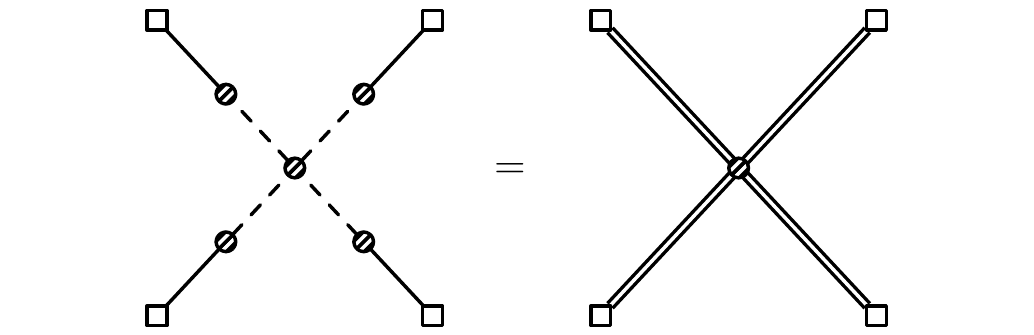}}
\ede
The expression for this diagram is very simple, and very similar to the one for bispectrum (\ref{3pt}),
\begin{align}
&~\la\de\phi(\tau,\mb k_1)\de\phi(\tau,\mb k_2)\de\phi(\tau,\mb k_3)\de\phi(\tau,\mb k_4)\ra_{\lam_4}'\n\\
=&~\FR{\pi^4\lam_2^4\lam_4^{}}{2048k_1^3k_2^3k_3^3}\,\text{Im}\int_0^\infty\FR{\di z}{z^4}I_+(\FR{k_1}{k_4}z)I_+(\FR{k_2}{k_4}z)I_+(\FR{k_3}{k_4}z)I_+(z).
\end{align}
The simplification here is even more significant, reducing 16 diagrams to 1.

On the other hand, we have contribution from two cubic self-interaction vertices of $\si$, at order $\lam_2^4\lam_3^2$, which is described by the following diagram,
\bge
\parbox{105mm}{\includegraphics{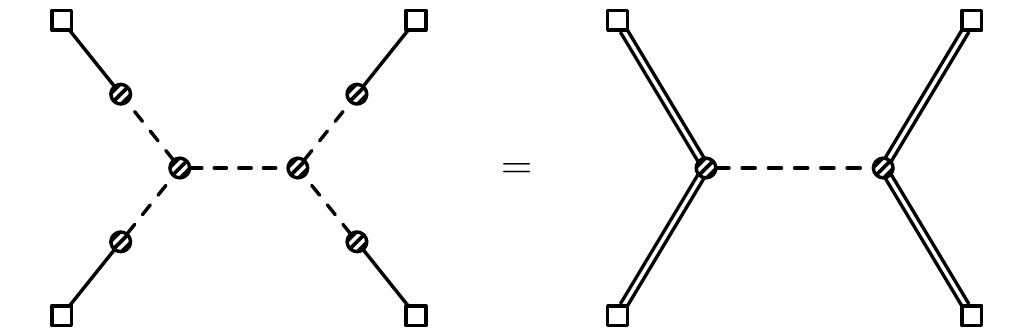}}
\ede
The above diagram shows the $s$-channel contribution only, and we should also include the corresponding $t$ and $u$-channels by simple permutations of external variables. The expression for the $s$-channel diagram can be written as,
\begin{align}
&\la\de\phi(\tau,\mb k_1)\de\phi(\tau,\mb k_2)\de\phi(\tau,\mb k_3)\de\phi(\tau,\mb k_4)\ra_{\lam_3^2,s}'\n\\
=&-\lam_2^4\sum_{a,b=\pm}\int_{-\infty}^0\FR{\di\tau_1\di\tau_2}{(H^2\tau_1\tau_2)^4} \mathcal{G}_a(k_1;\tau_1)\mathcal{G}_a(k_2;\tau_1)\mathcal{G}_b(k_3;\tau_2)\mathcal{G}_b(k_4;\tau_2)D_{ab}(k_I,\tau_1,\tau_2),
\end{align}
where $k_I=|\mb k_1+\mb k_2|$. We can compute this diagram in the same way as we did for bispectrum, namely, we firstly consider the 3-point function on the left side, which we denoted as $\mathcal{T}_\pm$, defined as follows,
\begin{align}
\mathcal{T}_+(k_1,k_2,k_I;\tau)
=&-\ii\lam_3\int_{-\infty}^0\FR{\di\tau_1}{(-H\tau_1)^4}\Big[\mathcal{G}_+(k_1;\tau_1)\mathcal{G}_+(k_2;\tau')D_{++}(k_I;\tau_1,\tau)\n\\
&~-\mathcal{G}_-(k_1;\tau_1)\mathcal{G}_-(k_2;\tau')D_{-+}(k_I;\tau_1,\tau)\Big].
\end{align}
It can be readily evaluated to be,
\begin{align}
\mathcal{T}_+(k_1,k_2,k_I;\tau)
=&-\FR{\ii\pi^3\lam_2^2\lam_3}{256k_1^3k_2^3}J_+(k_1/k_I,k_2/k_I;-k_I\tau),
\end{align}
where the function $J_+(k_1/k_I,k_2/k_I;z)$ reads,
\begin{align}
J_+(\FR{k_1}{k_I},\FR{k_2}{k_I};z)=&~e^{-\pi\,\text{Im}\,\nu}z^{3/2}\bigg\{2\,\text{Im}\bigg[H_{\nu}^{(1)}(z)\int_0^\infty\FR{\di z'}{z'^{5/2}}I_+(\FR{k_1}{k_I}z')I_+(\FR{k_2}{k_I}z')H_{\nu^*}^{(2)}(z')\bigg]\n\\
&~+H_{\nu^*}^{(2)}(z)\int_0^z\FR{\di z'}{z'^{5/2}}I_+(\FR{k_1}{k_I}z')I_+(\FR{k_2}{k_I}z')H_{\nu}^{(1)}(z')\n\\
&~-H_{\nu}^{(1)}(z)\int_0^z\FR{\di z'}{z'^{5/2}}I_+(\FR{k_1}{k_I}z')I_+(\FR{k_2}{k_I}z')H_{\nu^*}^{(2)}(z')\bigg\}.
\end{align}
Then, the bispectrum can be computed as,
\begin{align}
&\la\de\phi(\tau,\mb k_1)\de\phi(\tau,\mb k_2)\de\phi(\tau,\mb k_3)\de\phi(\tau,\mb k_4)\ra_{\lam_3^2,s}'\n\\
=&~2\lam_3\,\text{Im}\int_{-\infty}^0\FR{\di\tau}{(-H\tau)^4}\mathcal{T}_+(k_1,k_2,k_I;\tau)\mathcal{G}_+(k_3;\tau)\mathcal{G}_+(k_4;\tau)\n\\
=&-\FR{\pi^5\lam_2^4\lam_3^2}{8192H^2}\FR{k_I^3}{(k_1 k_2 k_3 k_4)^3}\,\text{Re}\int_0^{\infty}\FR{\di z}{z^4}J_+(\FR{k_1}{k_I},\FR{k_2}{k_I};z)I_+(\FR{k_3}{k_I}z)I_+(\FR{k_4}{k_I}z).
\end{align}
The above result should be able to be implemented into numerical calculations, and one may be able to learn more about the physics of trispectrum in quasi-single-field inflation. We shall continue the study of this topic elsewhere.

\section{Concluding Remarks}
\label{sec_Discussion}

In this paper, we introduced the Schwinger-Keldysh formalism in the language of path integral, adapted to the calculation of primordial perturbations in cosmology. We derived the diagrammatic rules for computing in-in correlation functions in a systematic and self-contained manner.

The SK path integral formulation and the diagrammatic rules derived from it are in principle equivalent to traditional canonical in-in formalism. However, the path-integral based diagrammatic method has multiple advantages, of which we list three here.

Firstly, it provides a visualized organizing principle for writing down the expressions of in-in correlation functions order by order in perturbation theory, which automatically takes care of the mechanical part of the calculation, such as perturbation expansion and Wick contraction.

Secondly, derivative couplings which appear frequently in cosmological context can be conveniently handled using path-integral based diagrammatics. We show in the two appendices of this paper that the diagrammatic rules can be derived directly from the classical Lagrangian, and that the in-in correlation functions calculated using this set of diagrammatic rules agree with canonical in-in formalism, even in the presence of derivative couplings.

Finally, this visualizable method allows us to understand the structure of the in-in amplitude better, and thus enable us to come up with tricks to simplify the calculation. We show in Sec.\;\ref{sec_QSFI} of this paper how to make dramatic simplification in the calculation of non-Gaussianities of quasi-single-field inflation by using the trick of mixed propagator.

Though for the most part of this paper we have the inflation background (namely quasi de Sitter space) in mind, the formalism described here should also apply to other cases with spatial homogeneity and isotropy but with nontrivial time evolution. In particular, it can be applied to other alternative-to-inflation scenarios, as well as general FRW universe in post-big-bang era.

The diagrammatic method we described here is most suitable for tree-level calculation, which should be able to bring some simplifications if one has to do everything numerically from the beginning (e.g., in the case of complicated time-dependence in the background solutions $\ob\phi^A(\tau)$).
On the other hand, we have not considered the loop corrections in this paper in detail. Due to the split of time and space coordinates, the loop calculation becomes rather tricky in this formalism. In the inflation background, the loop correction can sometimes be put into a dS covariant form. In such cases, the most suitable way to perform loop calculation is to use dS covariant techniques such as Wick rotation \cite{Chen:2016hrz}. But the loop calculations in more general settings remain challenging.

\paragraph{Acknowledgements.} We thank Andrew Cohen and Mohammad Hossein Namjoo for discussions.
XC is supported in part by the NSF grant PHY-1417421. YW is supported by grants HKUST4/CRF/13G, GRF 16301917 and ECS 26300316 issued by the Research Grants Council (RGC) of Hong Kong. ZZX is supported in part by Center of Mathematical Sciences and Applications, Harvard University.

\begin{appendix}

\section{Effective Lagrangian with Derivative Couplings}
\label{sec_DC}

In this appendix, we carry out the path integral over the momentum $\pi_\pm$ in (\ref{QevHPI}) for the theory with higher order derivative couplings. We shall show that the result of this integral is still given by (\ref{QevLPI}). When there is no derivative coupling, this has been shown in the main text, as can be seen from (\ref{ld_nd}), (\ref{hd_nd}), and (\ref{freepiint}). On the other hand, when there are higher order derivative couplings, we can only carry out the path integral over $\pi_\pm$ perturbatively. In this appendix, we verify (\ref{QevLPI}) to the 4th order in the power of fields. This is enough for a tree-level calculation of the trispectrum. The generalization to higher orders should be straightforward using our method described in this appendix, which we will consider in the future.

To 4th order in the power of fields, it is enough to consider the following classical Lagrangian,
\begin{align}
\label{Lcl}
\ld_\text{cl}
=&~\FR{1}{2}\mathcal{U}_{AB}\varphi'^A\varphi'^B+\mathcal{V}_A(\varphi)\varphi'^A+\mathcal{W}(\varphi)\n\\
&~+\FR{1}{2}\mathcal{X}_{AB}(\varphi)\varphi'^A\varphi'^B+\FR{1}{6}\mathcal{Y}_{ABC}(\varphi)\varphi'^A\varphi'^B\varphi'^C+\FR{1}{24}\mathcal{Z}_{ABCD}(\varphi)\varphi'^A\varphi'^B\varphi'^C\varphi'^D.
\end{align}
The first line is identical to (\ref{ld_nd}) which contains no derivative couplings, while the second line are derivative couplings, where the coefficients are totally symmetric in indices. The coefficient $\mathcal{X}_{AB}(\varphi)$ depends at least linearly on $\varphi^A$. Again, we use $\mathcal{U}_{AB}$ to lower the indices of fields and $\mathcal{U}^{AB}=(\mathcal{U}_{AB})^{-1}$ to raise indices.

We firstly derive the conjugate momentum $\pi_A$ and the Hamiltonian density $\hd$. The conjugate momentum is,
\bge
  \pi_A\equiv\FR{\pd\ld}{\pd\varphi'^A}=\big[\mathcal{U}+\mathcal{X}(\varphi)]_{AB}\varphi'^B+\mathcal{V}_A(\varphi)+\FR{1}{2}\mathcal{Y}_{ABC}(\varphi)\varphi'^B\varphi'^C+\FR{1}{6}\mathcal{Z}_{ABCD}(\varphi)\varphi'^B\varphi'^C\varphi'^D.
\ede
This time it is generally impossible to solve $\varphi'$ in terms of $\pi$ and $\varphi$. But we can treat the power of fields as a perturbation parameter, and solve $\varphi'$ perturbatively. Therefore, we define,
\begin{align}
\label{XYZexpand}
&\mathcal{X}_{AB}(\varphi)=\sum_{i\geq 1}\mathcal{X}_{AB}^{(i)}(\varphi),
&&\mathcal{Y}_{ABC}(\varphi)=\sum_{i\geq 0}\mathcal{Y}_{ABC}^{(i)}(\varphi),
&&\mathcal{Z}_{ABCD}(\varphi)=\sum_{i\geq 0}\mathcal{Z}_{ABCD}^{(i)}(\varphi),
\end{align}
where the superscript $(i)$ indicates that the corresponding term depends on $i$th power of $\varphi^A$. We don't need to expand $\mathcal{V}_A(\varphi)$ and $\mathcal{W}(\varphi)$ because they can be treated without perturbation expansion. Now, we are going to solve $\varphi'_A$ to 3rd order in fields, which is needed for a tree-level calculation of trispectrum\footnote{This is because we need to derive $\hd$ as well as $\ld_\text{eff}$ to 4th order in fields, while $\hd=\pi\varphi'-\ld_\text{cl}$ starts at 2nd order, so we only need to expand $\varphi'$ to 3rd order. A possible zeroth order term in $\mathcal{V}_A$ can be eliminated using integration-by-parts.}.
\bgs
\begin{align}
\varphi'^A=&\sum_{i=1}^3\varphi'^{(i)A}\\
\varphi'^{(1)A}=&~\wh\pi^A,\\
\varphi'^{(2)A}=&-\mathcal{X}^{(1)A}_{}{}_B^{}\wh\pi^{B}-\FR{1}{2}\mathcal{Y}^{(0)A}_{}{}_{BC}^{}\wh\pi^{B}\wh\pi^{C},\\
\varphi'^{(3)A}=&-\mathcal{X}^{(2)A}_{}{}_{B}^{}\wh\pi^B-\mathcal{X}^{(1)A}_{}{}_{B}^{}\varphi'^{(2)B}
-\FR{1}{2}\mathcal{Y}_{}^{(1)A}{}^{}_{BC}\wh\pi^B\wh\pi^C
-\mathcal{Y}_{}^{(0)A}{}^{}_{BC}\wh\pi^B\varphi'^{(2)C}\n\\
&-\FR{1}{6}\mathcal{Z}_{}^{(0)A}{}_{BCD}^{}\wh\pi^B\wh\pi^C\wh\pi^D,
\end{align}
\eds
where we have defined $\wh\pi^A\equiv(\pi-\mathcal{V})^A$. Then the Hamiltonian density can be derived to be,
\bgs
\label{Hpert}
\begin{align}
\hd=&\sum_{i=2}^{4}\hd^{(i)},\\
\hd^{(2)}
=&~\FR{1}{2}\wh\pi_A\wh\pi^A-\mathcal{W},\\
\hd^{(3)}
=&-\FR{1}{2}\mathcal{X}_{AB}^{(1)}\wh\pi^A\wh\pi^B-\FR{1}{6}\mathcal{Y}_{ABC}^{(0)}\wh\pi^A\wh\pi^B\wh\pi^C,\\
\label{Hpert4}
\hd^{(4)}
=&-\FR{1}{2}\Big(\mathcal{X}_{AB}^{(2)}-\mathcal{X}^{(1)}_{AC}\mathcal{X}^{(1)C}{}_B\Big)\wh\pi^A\wh\pi^B-\FR{1}{6}\Big(\mathcal{Y}^{(1)}_{ABC}-3\mathcal{X}_{AD}^{(1)}\mathcal{Y}^{(0)D}{}_{BC}\Big)\wh\pi^A\wh\pi^B\wh\pi^C\n\\
&-\FR{1}{24}\Big(\mathcal{Z}^{(0)}_{ABCD}-3\mathcal{Y}_{ABE}^{(0)}\mathcal{Y}^{(0)E}{}_{CD}\Big)\wh\pi^A\wh\pi^B\wh\pi^C\wh\pi^D,
\end{align}
\eds
Clearly, the 2nd order term $\hd^{(2)}$ here recovers the Hamiltonian (\ref{hd_nd}) found for non-derivative coupling theories. Furthermore, as a byproduct, if we substitute $\pi^A\to\varphi'^A$ in above expression, we get the Hamiltonian density $\hd_I$ in interaction picture which is useful for the canonical in-in formalism in the operator language, see (\ref{Hint}). It is worth noting that the coupling terms in the Hamiltonian in interaction picture is not a simple sign reverse of the classical Lagrangian due to the derivative couplings, and the nontrivial terms appear starting from the 4th order in (\ref{Hpert4}). This complicates the canonical in-in formalism. Now we are going to show that these additional terms are actually canceled in the Lagrangian appeared in the path integral (\ref{QevLPI}), and therefore the diagrammatic rules follow directly from the classical Lagrangian even with the presence of arbitrary derivative couplings.

We do the path integral of momentum for $\pi_{+A}$ in (\ref{QevHPI}) to 4th order, and the method for doing integral over $\pi_{-A}$ is completely the same. Therefore we will drop the $+$ sign from now on for clarity. To do the path integral (\ref{QevHPI}) for the Hamiltonian (\ref{Hpert}), It is convenient to shift the momentum variable in (\ref{QevHPI}) from $\pi_A$ to $\wh\pi_A=\pi_A-\mathcal{V}_A$, and introduce an external source $K^A$ for $\wh\pi_A$, so that the path integral over $\pi_A$ becomes,
\begin{align}
\label{IK}
I[K]
=&\int\mathcal{D}\wh\pi\exp\bigg[\ii\int_{\tau_0}^{\tau_f}\di\tau\di^3\mb x\,\Big(\pi_{A}\varphi'^{A}+\wh\pi_A K^{A}-{\hd}[\wh\pi,\varphi]\Big)\bigg]\n\\
=&\int\mathcal{D}\wh\pi\exp\bigg[\ii\int_{\tau_0}^{\tau_f}\di\tau\di^3\mb x\,\Big(\wh\pi_{A}\big(\varphi'+K\big)^{A}+\mathcal{V}_A\varphi'^{A}-{\hd}^{(2)}[\wh\pi,\varphi]-\wt{\hd}[\wh\pi,\varphi]\Big)\bigg],
\end{align}
where we have defined $\wt\hd[\wh\pi,\varphi]$ to be the higher order derivative coupling terms in the Hamiltonian, which are given by $\hd^{(3)}+\hd^{(4)}$ from (\ref{Hpert}) in our case. Then as usual, we can turn $\wt\hd$ into functional derivative, as,
\begin{align}
I[K]
=&\,\exp\bigg(-\ii\int_{\tau_0}^{\tau_f}\di\tau\di^3\mb x\,\wt{\hd}\Big[\FR{\de}{\ii\de K},\varphi\Big]\bigg)\n\\
&~\times\int\mathcal{D}\wh\pi\exp\bigg[\ii\int_{\tau_0}^{\tau_f}\di\tau\di^3\mb x\,\Big(\wh\pi_{A}\big(\varphi'+K\big)^{A}+\mathcal{V}_A\varphi'^{A}-{\hd}^{(2)}[\wh\pi,\varphi]\Big)\bigg]\n\\
=&~\exp\bigg(-\ii\int_{\tau_0}^{\tau_f}\di\tau\di^3\mb x\,\wt{\hd}\Big[\FR{\de}{\ii\de K},\phi_+;\tau\Big]\bigg)\n\\
&\times\exp\bigg[\ii\int_{\tau_0}^{\tau_f}\di\tau\di^3\mb x\,\Big(\FR{1}{2}(\varphi'+K)_A(\varphi'+K)^A+\mathcal{V}_A\varphi'^A+\mathcal{W}\Big)\bigg].
\end{align}
Then we define an effective action $S_{\text{eff}}$, as well as the corresponding effective Lagrangian $\ld_{\text{eff}}$, as,
\bge
\label{Seff}
  S_{\text{eff}}=\int\di\tau\di^3\mb x\,\ld_{\text{eff}}\equiv -\ii \log I[K=0],
\ede
and therefore our task is to show $\ld_\text{eff}=\ld_\text{cl}$ perturbatively. To the 4th order in the power of fields, the effective Lagrangian $\ld_\text{eff}$ is given by the following functional derivatives,
\begin{align}
\label{LeffFD}
\ld_{\text{eff}}
=&-\ii\log\bigg\{\bigg[1-\ii\int\di^4x\,\Big(\hd^{(3)}\Big[\FR{\de}{\ii\de K},\varphi\Big]+\hd^{(4)}\Big[\FR{\de}{\ii\de K},\varphi\Big]\Big)-\FR{1}{2}\Big(\int\di^4x\,\hd^{(3)}\Big[\FR{\de}{\ii\de K},\varphi\Big]\Big)^2\bigg]\n\\
&~\times\exp\Big[\ii\int\di^4x\,\Big(\FR{1}{2}(\varphi'+K)_A(\varphi'+K)^A+\mathcal{V}_A\varphi'^A+\mathcal{W}\Big)\Big]\bigg\}_{K=0}.
\end{align}
The functional derivative can be carried out most easily by fictitious Feynman diagrams, where we can think of $\varphi'^A$ as external sources, and $\mathcal{U}^{AB}$ as internal propagator. The external legs make no contribution, and the vertices can be directly read from $\wt\hd$. Consequently, the interaction terms introduced by functional derivatives can be represented by the following diagrams. At 2nd order in $\varphi'^A$,
\bge
\parbox{120mm}{\includegraphics{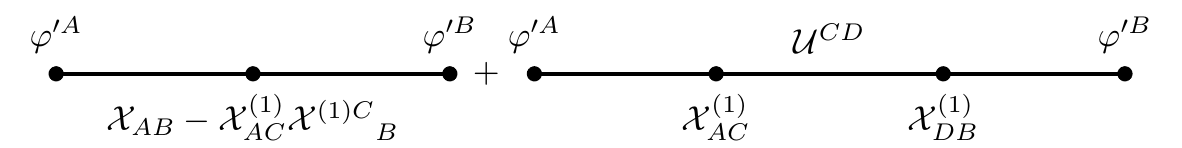}}
\ede
The contribution from these two diagrams are,
\begin{align}
\ld_\text{eff}\supset&~ \FR{1}{2}\Big(\mathcal{X}_{AB}^{(1)}+\mathcal{X}_{AB}^{(2)}-\mathcal{X}_{AC}^{(1)}\mathcal{X}^{(1)C}{}_B^{}\Big)\varphi'^A\varphi'^B+\FR{1}{2}\mathcal{X}_{AC}^{(1)}\mathcal{X}_{DB}^{(1)}\mathcal{U}^{CD}\varphi'^A\varphi'^B\n\\
=&~\FR{1}{2}\Big(\mathcal{X}_{AB}^{(1)}+\mathcal{X}_{AB}^{(2)}\Big)\varphi'^A\varphi'^B.
\end{align}
Then at 3rd order in $\varphi'^A$,
\bge
\parbox{140mm}{\includegraphics{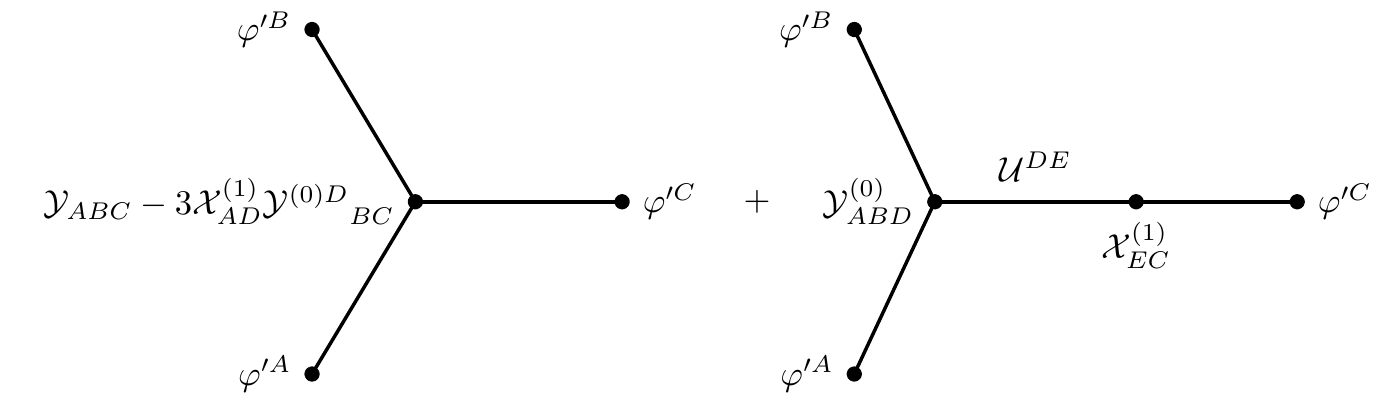}}
\ede
and there are two additional permutations of the second diagram where the $\mathcal{X}$-vertex can appear in the other two legs. All these diagrams contribute to $\ld_\text{eff}$ as,
\begin{align}
\ld_\text{eff}\supset&~
\FR{1}{6}\Big(\mathcal{Y}^{(0)}_{ABC}+\mathcal{Y}^{(1)}_{ABC}-3\mathcal{X}_{AD}^{(1)}\mathcal{Y}^{(0)D}{}_{BC}^{}\Big)\varphi'^A\varphi'^B\varphi'^C+3\times\FR{1}{6}\mathcal{X}_{EC}^{(1)}\mathcal{Y}_{ABD}^{(0)}\mathcal{U}^{DE}\varphi'^A\varphi'^B\varphi'^C\n\\
=&~\FR{1}{6}\Big(\mathcal{Y}^{(0)}_{ABC}+\mathcal{Y}^{(1)}_{ABC}\Big)\varphi'^A\varphi'^B\varphi'^C
\end{align}

Finally, at 4rd order in $\varphi'^A$,
\bge
\parbox{140mm}{\includegraphics{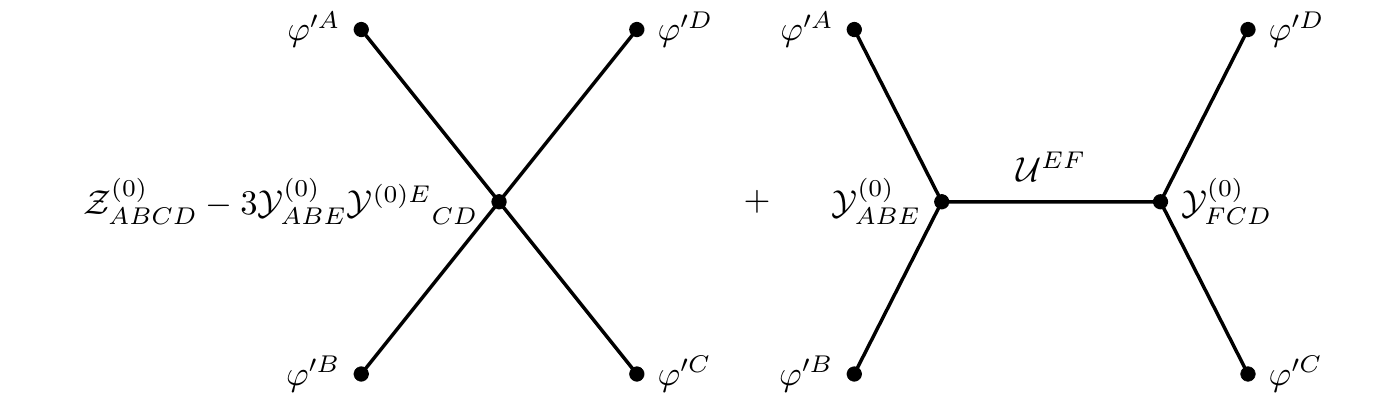}}
\ede
and again we have two additional permutations of the second ($s$-channel) diagram representing the $t$-channel and $u$-channel contributions, respectively. The contribution from these four diagrams is,
\begin{align}
\ld_\text{eff}\supset&~
\FR{1}{24}\Big(\mathcal{Z}_{ABCD}^{(0)}-3\mathcal{Y}_{ABE}^{(0)}\mathcal{Y}^{(0)E}{}_{CD}^{}\Big)\varphi'^A\varphi'^B\varphi'^C\varphi'^D+3\times\FR{1}{24}\mathcal{Y}_{ABE}^{(0)}\mathcal{Y}_{FCD}^{(0)}\mathcal{U}^{EF}\varphi'^A\varphi'^B\varphi'^C\varphi'^D\n\\
=&~\FR{1}{24}\mathcal{Z}_{ABCD}^{(0)}\varphi'^A\varphi'^B\varphi'^C\varphi'^D.
\end{align}
Summing over all diagrams, and also including the terms in the second line of (\ref{LeffFD}), we see that the additional terms in the Hamiltonian density (\ref{Hpert4}) are completely canceled out in the effective Lagrangian, and thus we have verified that $\ld_\text{eff}=\ld_\text{cl}$ holds to 4th order in the number of fields.

{
In literature there is also a more conventional approach to this problem in which one expands the Lagrangian in power series of derivative of fields $\varphi'$ and keeps non-derivative fields $\varphi$ to all orders. In this approach, one can no longer expand the field-dependent coefficients such as $\mathcal X_{AB}(\varphi)$ in power series of fields as we did in (\ref{XYZexpand}). In particular, the momentum integral would generate a nontrivial field-dependent determinant $\det[\mathcal{U}+\mathcal{X}(\varphi)]$ and this determinant would make additional contribution to Feynman rules which cannot be read off directly from the classical Lagrangian. We clarify that our approach is more modest compared with this conventional approach in that we do not seek for an expression to all orders in power of fields, but only to finite orders, because a Lagrangian to finite power of fields is all we need to calculate tree-level non-Gaussianity with given number of external legs.

}

\section{Sample Calculation with Derivative Couplings}
\label{sec_SampleDC}

In this appendix we demonstrate the diagrammatic calculation in the theory with derivative couplings, by consider the tree-level trispectrum contributed by a 3rd order derivative coupling. In canonical in-in formalism, there is an additional 4th order term in the Hamiltonian in interaction picture, cf. (\ref{Hpert4}), which is absent in the effective Lagrangian of the diagrammatic approach. The point of this appendix is to show how the additional 4th order contribution in canonical in-in formalism is automatically taken into account using path-integral based diagrammatic method.

To be specific, we consider the following interaction,
\bge
  S\supset\FR{y\dot\phi_0}{6}\int\di\tau\di^3\mb x\,a(\tau){\de\phi}'^3,
\ede
We also assume that the free part of the Hamiltonian for $\de\phi$ has the following form,
\bge
  \hd^{(2)}[\de\phi',\de\phi]=\FR{1}{2}c(\tau)\de\phi'^2+\cdots,
\ede
where we assume the coefficient of $\phi'^2$ term to be an unspecified function $c(\tau)$, while the rest of terms are unimportant. Then, the mode function for $\de\phi$ is normalized according to,
\bge
\label{norm}
  c(\tau)\big[u_k(\tau)u_k^{*}{'}(\tau)-u_k^*(\tau)u_k'(\tau)\big]=\ii.
\ede
Therefore, the model we are considering here is identical to (\ref{Hpert}), but with only one real scalar $\de\phi$ and one nonzero interaction term $\mathcal{Y}_{ABC}^{(0)}=a(\tau)y\dot\phi_0$. Furthermore, the ``metric'' $\mathcal{U}_{AB}=c(\tau)$.

\paragraph{Canonical approach.} We firstly compute the 4-point function using canonical in-in formalism. For this purpose we need to compute the Hamiltonian in interaction picture. This can be read directly from (\ref{Hpert}),
\bge
\label{HIsample}
  \hd_I=\hd^{(2)}[{\de\phi}',\de\phi]-\FR{1}{6}y\dot\phi_0a(\tau){\de\phi'}^3-\FR{1}{8}y^2\dot\phi_0^2c^{-1}(\tau)a^2(\tau){\de\phi'}^4,
\ede
where $\hd^{(2)}$ is the free part of the Hamiltonian, and the second term proportional to $y$ is the direct sign-flip of the Lagrangian, while the last term proportional to $y^2$ is new.

Therefore we need to compute two types of ``diagrams'', corresponding to the second and third terms in (\ref{HIsample}), as shown below,
\bge
\parbox{105mm}{\includegraphics{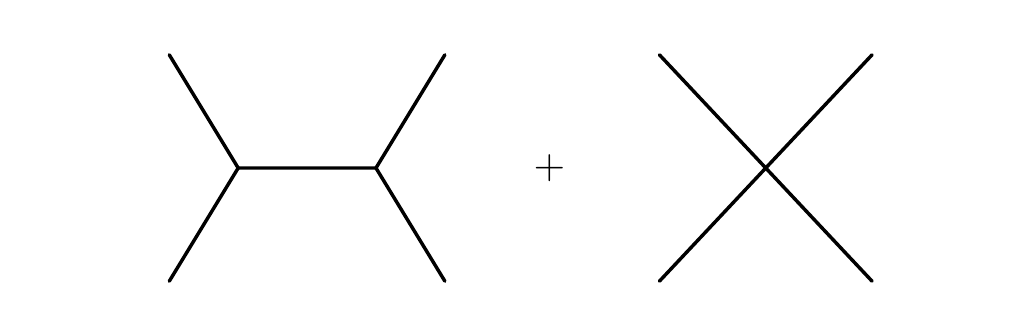}}
\ede
There are also two similar permutations to the first diagram corresponding to $t$-channel and $u$-channel. But we only display $s$-channel explicitly. This $s$-channel diagram reads,
\begin{align}
\label{4ptOper1}
&-2y^2\dot\phi_0^2\,\text{Re}\int_{-\infty}^0\di\tau_1\di\tau_2\, a(\tau_1)a(\tau_2)\Big\{
\big[\pd_{\tau_1}G_<(k_1;\tau_1,0)\big]
\big[\pd_{\tau_1}G_<(k_2;\tau_1,0)\big]
\big[\pd_{\tau_2}G_<(k_3;\tau_2,0)\big]\n\\&~\times\big[\pd_{\tau_2}G_<(k_4;\tau_2,0)\big]
\Big[\theta(\tau_1-\tau_2)\pd_{\tau_1}\pd_{\tau_2}G_>(k_S;\tau_1,\tau_2)+\theta(\tau_2-\tau_1)\pd_{\tau_1}\pd_{\tau_2}G_<(k_S;\tau_1,\tau_2)\Big]\n\\
&~+\big[\pd_{\tau_1}G_<(k_1;\tau_1,0)\big]
\big[\pd_{\tau_1}G_<(k_2;\tau_1,0)\big]
\big[\pd_{\tau_2}G_>(k_3;\tau_2,0)\big]\big[\pd_{\tau_2}G_>(k_4;\tau_2,0)\big]
\pd_{\tau_1}\pd_{\tau_2}G_<(k_S;\tau_1,\tau_2)
\Big\},
\end{align}
where we have used (\ref{Ggs}) to represent all mode functions for clarity, and $k_S=|\mb k_1+\mb k_2|$. The contributions from $t$-channel and $u$-channel can be got by $k_1\leftrightarrow k_4$ and $k_1\leftrightarrow k_3$, respectively. On the other hand, the second diagram with 4-point interaction reads,
\begin{align}
\label{4ptOper2}
6y^2\dot\phi_0^2\,\text{Im}\int_{-\infty}^0\di\tau\,\FR{a^2(\tau)}{c(\tau)}
\big[\pd_{\tau}G_>(k_1;\tau,0)\big]
\big[\pd_{\tau}G_>(k_2;\tau,0)\big]
\big[\pd_{\tau}G_>(k_3;\tau,0)\big]
\big[\pd_{\tau}G_>(k_4;\tau,0)\big].
\end{align}
Then the 4-point function of order $y^2$ is the sum of the above 4 diagrams.

\paragraph{Diagrammatic approach.} Next we compute the same 4-point function using diagrammatic method. According to the diagrammatic rules, there is only one type of diagram,
\bge
\parbox{105mm}{\includegraphics{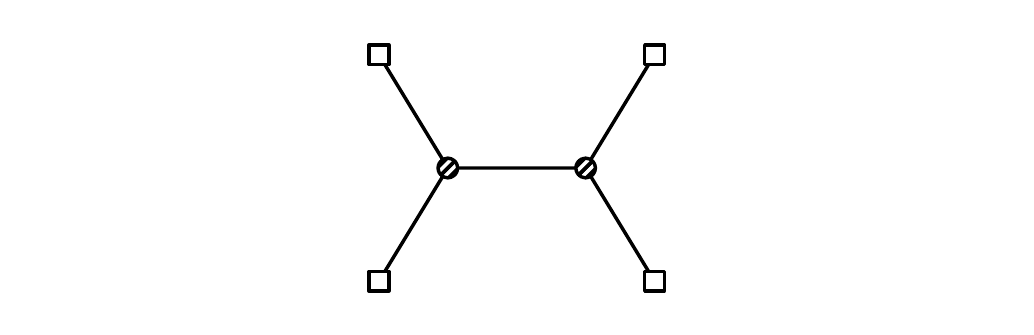}}
\ede
as well as corresponding $t$-channel and $u$-channel. This $s$-channel diagram can be written as,
\begin{align}
\label{4ptD}
&\la\phi(\tau,\mb k_1)\phi(\tau,\mb k_2)\phi(\tau,\mb k_3)\phi(\tau,\mb k_4)\ra_{y^2}'\n\\
=&~y^2\dot\phi_0^2\sum_{a,b=\pm}\int_{\tau_0}^{\tau}\di\tau_1\di\tau_1 a(\tau_1) a(\tau_2)
\big[\pd_{\tau_1}G_{a}(k_1;\tau_1)\big]
\big[\pd_{\tau_1}G_{a}(k_2;\tau_1)\big]\n\\
&~\times\big[\pd_{\tau_2}G_{b}(k_3;\tau_2)\big]
\big[\pd_{\tau_2}G_{b}(k_4;\tau_2)\big]
\pd_{\tau_1}\pd_{\tau_2}G_{ab}(k_S;\tau_1,\tau_2),
\end{align}
where $k_S=|\mb k_1+\mb k_2|$. To see this expression indeed equals to the result found in canonical approach, namely the sum of (\ref{4ptOper1}) and (\ref{4ptOper2}) (together with $(t,u)$-channels), we only need to use the following identity,
\begin{align}
\label{ddGpp}
\pd_{\tau_1}\pd_{\tau_2}G_{++}(k;\tau_1,\tau_2)=&~\theta(\tau_1-\tau_2)\pd_{\tau_1}\pd_{\tau_2}G_>(k;\tau_1,\tau_2)+\theta(\tau_2-\tau_1)\pd_{\tau_1}\pd_{\tau_2}G_<(k;\tau_1,\tau_2)\n\\
&~+\ii c^{-1}(\tau)\de(\tau_1-\tau_2).
\end{align}
Similar identities can be written down for $G_{--}$. The important point here is that if we apply two time derivatives on the propagator $G_{\pm\pm}(k;\tau_1,\tau_2)$, there will be an ``non-covariant'' contact term proportional to $\de(\tau_1-\tau_2)$. It is this contact term that generates the contribution identical to (\ref{4ptOper2}), while the contribution from the rest of terms correspond to (\ref{4ptOper1}). As an explicit check, we substitute the $\de$-function term in (\ref{ddGpp}) back to (\ref{4ptD}), and get,
\bge
  2y^2\dot\phi_0^2\,\text{Im}\int_{-\infty}^0\di\tau\,\FR{a^2(\tau)}{c(\tau)}
\big[\pd_{\tau}G_>(k_1;\tau,0)\big]
\big[\pd_{\tau}G_>(k_2;\tau,0)\big]
\big[\pd_{\tau}G_>(k_3;\tau,0)\big]
\big[\pd_{\tau}G_>(k_4;\tau,0)\big].
\ede
Taking $(t,u)$-channels into account, we have another factor $3$, and thus the result indeed matches (\ref{4ptOper2}) found in operator formalism. Therefore we have shown that the 4-point functions calculated using canonical method and path-integral based diagrammatic method agree with each other.

At the end of this appendix, we prove the identity (\ref{ddGpp}), as follows,
\begin{align}
&\int_{\tau_0}^{\tau_f}\di\tau_1\, f(\tau_1)\int_{\tau_0}^{\tau_f}\di\tau_2\,g(\tau_2) \pd_{\tau_1}\pd_{\tau_2}G_{++}(k;\tau_1,\tau_2)\n\\
=&\int_{\tau_0}^{\tau_f}\di\tau_1\, f(\tau_1)\int_{\tau_0}^{\tau_f}\di\tau_2\,g(\tau_2) \Big[\theta(\tau_1-\tau_2)\pd_{\tau_1}\pd_{\tau_2}G_{>}+\theta(\tau_2-\tau_1)\pd_{\tau_1}\pd_{\tau_2}G_{<}\n\\
&~-\de(\tau_1-\tau_2)(\pd_{\tau_1}-\pd_{\tau_2})(G_>-G_<)-(G_>-G_<)\pd_{\tau_1}\de(\tau_1-\tau_2)\Big]\n\\
=&\int_{\tau_0}^{\tau_f}\di\tau_1\, f(\tau_1)\int_{\tau_0}^{\tau_f}\di\tau_2\,g(\tau_2) \Big[\theta(\tau_1-\tau_2)\pd_{\tau_1}\pd_{\tau_2}G_{>}+\theta(\tau_2-\tau_1)\pd_{\tau_1}\pd_{\tau_2}G_{<}\n\\
&~+\Big(u(\tau_1)u^*{}'(\tau_2)-u^*(\tau_1)u'(\tau_2)\Big)\de(\tau_1-\tau_2)\n\\
&~+\FR{f'(\tau_1)}{f(\tau_1)}g(\tau_2)\Big(u(\tau_1)u(\tau_2)^*-u(\tau_1)^*u(\tau_2)\Big)\de(\tau_1-\tau_2)\Big]\n\\
=&\int_{\tau_0}^{\tau_f}\di\tau_1\, f(\tau_1)\int_{\tau_0}^{\tau_f}\di\tau_2\,g(\tau_2)\Big[\theta(\tau_1-\tau_2)\pd_{\tau_1}\pd_{\tau_2}G_{>}+\theta(\tau_2-\tau_1)\pd_{\tau_1}\pd_{\tau_2}G_{<}\Big]\n\\
&~+\ii\int_{\tau_0}^{\tau_f}\di\tau_1\,\FR{f(\tau_1)g(\tau_1)}{c(\tau_1)},
\end{align}
where $f(\tau)$ and $g(\tau)$ are arbitrary functions, and we have used the normalization condition (\ref{norm}).

\end{appendix}


\begin{thebibliography}{999}



\bibitem{Chen:2010xka}
  X.~Chen,
  ``Primordial Non-Gaussianities from Inflation Models,''
  Adv.\ Astron.\  {\bf 2010} (2010) 638979
  [arXiv:1002.1416 [astro-ph.CO]].

\bibitem{Wang:2013eqj}
  Y.~Wang,
  ``Inflation, Cosmic Perturbations and Non-Gaussianities,''
  Commun.\ Theor.\ Phys.\  {\bf 62} (2014) 109
  [arXiv:1303.1523 [hep-th]].



\bibitem{Schwinger:1960qe}
  J.~S.~Schwinger,
  ``Brownian motion of a quantum oscillator,''
  J.\ Math.\ Phys.\  {\bf 2} (1961) 407.


\bibitem{Keldysh:1964ud}
  L.~V.~Keldysh,
  ``Diagram technique for nonequilibrium processes,''
  Zh.\ Eksp.\ Teor.\ Fiz.\  {\bf 47} (1964) 1515
   [Sov.\ Phys.\ JETP {\bf 20} (1965) 1018].

\bibitem{Feynman:1963fq}
  R.~P.~Feynman and F.~L.~Vernon, Jr.,
  ``The Theory of a general quantum system interacting with a linear dissipative system,''
  Annals Phys.\  {\bf 24} (1963) 118.

\bibitem{Landau10}
  L.~D.~Landau, E.~M.~Lifshitz and L.~P.~Pitaevskij,
  \emph{Course of Theoretical Physics. Vol. 10: Physical Kinetics}.
  Butterworth-Heinemann, 1981.

\bibitem{Chou:1984es}
  K.~c.~Chou, Z.~b.~Su, B.~l.~Hao and L.~Yu,
  ``Equilibrium and Nonequilibrium Formalisms Made Unified,''
  Phys.\ Rept.\  {\bf 118} (1985) 1.

\bibitem{Jordan:1986ug}
  R.~D.~Jordan,
  ``Effective Field Equations for Expectation Values,''
  Phys.\ Rev.\ D {\bf 33} (1986) 444.

\bibitem{Haehl:2016pec}
  F.~M.~Haehl, R.~Loganayagam and M.~Rangamani,
  ``Schwinger-Keldysh formalism I: BRST symmetries and superspace,''
  arXiv:1610.01940 [hep-th].

\bibitem{Tsamis:1996qq}
  N.~C.~Tsamis and R.~P.~Woodard,
  ``Quantum gravity slows inflation,''
  Nucl.\ Phys.\ B {\bf 474} (1996) 235
  [hep-ph/9602315].

\bibitem{Tsamis:1996qm}
  N.~C.~Tsamis and R.~P.~Woodard,
  ``The Quantum gravitational back reaction on inflation,''
  Annals Phys.\  {\bf 253} (1997) 1
  [hep-ph/9602316].

\bibitem{Seery:2007we}
  D.~Seery,
  ``One-loop corrections to a scalar field during inflation,''
  JCAP {\bf 0711} (2007) 025
  [arXiv:0707.3377 [astro-ph]].

\bibitem{vanderMeulen:2007ah}
  M.~van der Meulen and J.~Smit,
  ``Classical approximation to quantum cosmological correlations,''
  JCAP {\bf 0711} (2007) 023
  [arXiv:0707.0842 [hep-th]].

\bibitem{Seery:2008ax}
  D.~Seery, M.~S.~Sloth and F.~Vernizzi,
  ``Inflationary trispectrum from graviton exchange,''
  JCAP {\bf 0903} (2009) 018
  [arXiv:0811.3934 [astro-ph]].

\bibitem{Leblond:2010yq}
  L.~Leblond and E.~Pajer,
  ``Resonant Trispectrum and a Dozen More Primordial N-point functions,''
  JCAP {\bf 1101} (2011) 035
  [arXiv:1010.4565 [hep-th]].


\bibitem{Chen:2016nrs}
  X.~Chen, Y.~Wang and Z.~Z.~Xianyu,
  ``Loop Corrections to Standard Model Fields in Inflation,''
  JHEP {\bf 1608} (2016) 051
  [arXiv:1604.07841 [hep-th]].

\bibitem{Calzetta:1986ey}
  E.~Calzetta and B.~L.~Hu,
  ``Closed Time Path Functional Formalism in Curved Space-Time: Application to Cosmological Back Reaction Problems,''
  Phys.\ Rev.\ D {\bf 35} (1987) 495.

\bibitem{Weinberg:2005vy}
  S.~Weinberg,
  ``Quantum contributions to cosmological correlations,''
  Phys.\ Rev.\ D {\bf 72} (2005) 043514
  [hep-th/0506236].

\bibitem{Prokopec:2010be}
  T.~Prokopec and G.~Rigopoulos,
  ``Path Integral for Inflationary Perturbations,''
  Phys.\ Rev.\ D {\bf 82} (2010) 023529
  [arXiv:1004.0882 [gr-qc]].

\bibitem{Gong:2016qpq}
  J.~O.~Gong, M.~S.~Seo and G.~Shiu,
  ``Path integral for multi-field inflation,''
  JHEP {\bf 1607} (2016) 099
  [arXiv:1603.03689 [hep-th]].


\bibitem{Chen:2009we}
  X.~Chen and Y.~Wang,
  ``Large non-Gaussianities with Intermediate Shapes from Quasi-Single Field Inflation,''
  Phys.\ Rev.\ D {\bf 81}, 063511 (2010)
  [arXiv:0909.0496 [astro-ph.CO]].


\bibitem{Chen:2009zp}
  X.~Chen and Y.~Wang,
  ``Quasi-Single Field Inflation and Non-Gaussianities,''
  JCAP {\bf 1004} (2010) 027
  [arXiv:0911.3380 [hep-th]].


\bibitem{Baumann:2011nk}
  D.~Baumann and D.~Green,
  ``Signatures of Supersymmetry from the Early Universe,''
  Phys.\ Rev.\ D {\bf 85}, 103520 (2012)
  [arXiv:1109.0292 [hep-th]].

\bibitem{Assassi:2012zq}
  V.~Assassi, D.~Baumann and D.~Green,
  ``On Soft Limits of Inflationary Correlation Functions,''
  JCAP {\bf 1211}, 047 (2012)
  [arXiv:1204.4207 [hep-th]].


\bibitem{Chen:2012ge}
  X.~Chen and Y.~Wang,
  ``Quasi-Single Field Inflation with Large Mass,''
  JCAP {\bf 1209} (2012) 021
  [arXiv:1205.0160 [hep-th]].



\bibitem{Noumi:2012vr}
  T.~Noumi, M.~Yamaguchi and D.~Yokoyama,
  ``Effective field theory approach to quasi-single field inflation and effects of heavy fields,''
  JHEP {\bf 1306}, 051 (2013)
  [arXiv:1211.1624 [hep-th]].

\bibitem{Arkani-Hamed:2015bza}
  N.~Arkani-Hamed and J.~Maldacena,
  ``Cosmological Collider Physics,''
  arXiv:1503.08043 [hep-th].

\bibitem{Gong:2013sma}
  J.~O.~Gong, S.~Pi and M.~Sasaki,
  ``Equilateral non-Gaussianity from heavy fields,''
  JCAP {\bf 1311}, 043 (2013)
  [arXiv:1306.3691 [hep-th]].


\bibitem{Emami:2013lma}
  R.~Emami,
  JCAP {\bf 1404}, 031 (2014)
  [arXiv:1311.0184 [hep-th]].

\bibitem{Kehagias:2015jha}
  A.~Kehagias and A.~Riotto,
  ``High Energy Physics Signatures from Inflation and Conformal Symmetry of de Sitter,''
  Fortsch.\ Phys.\  {\bf 63}, 531 (2015)
  [arXiv:1501.03515 [hep-th]].

\bibitem{Chen:2015lza}
  X.~Chen, M.~H.~Namjoo and Y.~Wang,
  ``Quantum Primordial Standard Clocks,''
  JCAP {\bf 1602}, no. 02, 013 (2016)
  [arXiv:1509.03930 [astro-ph.CO]].


\bibitem{Bonga:2015urq}
  B.~Bonga, S.~Brahma, A.~S.~Deutsch and S.~Shandera,
  JCAP {\bf 1605}, no. 05, 018 (2016)
  [arXiv:1512.05365 [astro-ph.CO]].

\bibitem{Lee:2016vti}
  H.~Lee, D.~Baumann and G.~L.~Pimentel,
  ``Non-Gaussianity as a Particle Detector,''
  JHEP {\bf 1612} (2016) 040
  [arXiv:1607.03735 [hep-th]].



\bibitem{WeinbergQFTI}
  S.~Weinberg,
  \emph{The Quantum theory of fields. Vol. 1: Foundations},
  Cambridge, 1995.
  
\bibitem{Huang:2006eha}
  X.~Chen, M.~x.~Huang and G.~Shiu,
  ``The Inflationary Trispectrum for Models with Large Non-Gaussianities,''
  Phys.\ Rev.\ D {\bf 74} (2006) 121301
  [hep-th/0610235].


\bibitem{Chen:2016hrz}
  X.~Chen, Y.~Wang and Z.~Z.~Xianyu,
  ``Standard Model Mass Spectrum in Inflationary Universe,''
  arXiv:1612.08122 [hep-th].





\bibitem{Creminelli:2005hu}
  P.~Creminelli, A.~Nicolis, L.~Senatore, M.~Tegmark and M.~Zaldarriaga,
  ``Limits on non-gaussianities from wmap data,''
  JCAP {\bf 0605}, 004 (2006)
  [astro-ph/0509029].


\bibitem{Chen:2006nt}
  X.~Chen, M.~x.~Huang, S.~Kachru and G.~Shiu,
  ``Observational signatures and non-Gaussianities of general single field inflation,''
  JCAP {\bf 0701}, 002 (2007)
  [hep-th/0605045].



\bibitem{Meerburg:2016zdz}
  P.~D.~Meerburg, M.~M\"unchmeyer, J.~B.~Mu\~noz and X.~Chen,
  ``Prospects for Cosmological Collider Physics,''
  arXiv:1610.06559 [astro-ph.CO].


\end{thebibliography}
\end{document}